\documentclass[preprint,3p]{elsarticle}

\usepackage{amssymb}
\usepackage{bm}
\usepackage{amsmath}
\usepackage{caption}
\usepackage{cleveref}
\usepackage{multirow}
\newcommand{\degree}{^\circ}
\begin{document}
\captionsetup[figure]{labelfont={bf},name={Fig.},labelsep=period}
\captionsetup[table]{labelfont={bf},name={Table.},labelsep=period}

\begin{frontmatter}

\title{An accelerated sharp-interface method for multiphase flows simulations}

\author{Tian Long}%\corref{cor1}\fnref{label3}}
%\cortext[cor1]{I am corresponding author}
%\fntext[label3]{I also want to inform about\ldots}
%\fntext[label4]{Small city}

%\ead{author.one@mail.com}
%\ead[url]{author-one-homepage.com}
%address[label1]{Some University}
\author{Jinsheng Cai}
\author{Shucheng Pan\corref{cor1}}
\ead{shucheng.pan@nwpu.edu.cn}
\cortext[cor1]{Corresponding author}
%\ead{author.three@mail.com}
\address{School of Aeronautics, Northwestern Polytechnical University, Xi'an, 710072, PR China}

\begin{abstract}
 In this work, we develop an accelerated sharp-interface method based on (Hu et al., JCP, 2006) and (Luo et al., JCP, 2015) for multiphase flows simulations. Traditional multiphase simulation methods use the minimum time step of all fluids obtained according to CFL conditions to evolve the fluid states, which limits the computational efficiency, as the sound speed $c$ of one fluid may be much larger than the others. To address this issue, based on the original GFM-like sharp interface methods, the present method is developed by solving the  governing equations of each individual fluid with the corresponding time step. Without violating the numerical stability requirement, the states of fluid with larger time-scale features will be updated with a larger time step. The interaction step between two fluids is solved for synchronization, which is handled by interpolating the intermediate states of fluid with larger time-scale features. In addition, an interfacial flux correction is implemented to maintain the conservative property. The present method can be combined with a wavelet-based adaptive multi-resolution algorithm (Han et al., JCP, 2014) to achieve additional computational efficiency. A number of numerical tests indicate that the accuracy of the results obtained by the present method is comparable to the original costly method, with a significant speedup.
\end{abstract}


\end{frontmatter}

% \linenumbers

% main text
\section{Introduction}
\label{sec1}
%\subsection{Sample subsection}
%\label{subsec1}
Nowadays, the study of multiphase flows has attracted much attention from academia due to its widely practical usage in many branches of industry such as the bubble column reactors \cite{Olmos2001Numerical}, the solidification process of metals \cite{Xiong2006Multiphase}, the civitation in the naval industry \cite{Bonfiglio2016multiphase}, etc. The main challenge in the numerical simulation of multiphase flows is the treatment of the material interface which separates interacting different fluids. 

In general, the approaches for modelling interface interactions can be divided into two classes, the diffuse-interface methods and the sharp-interface methods. In the diffuse-interface methods, with limited thickness, the interface is often defined as an artificial region where the transition from one phase to another is smooth. Among those diffuse-interface methods, volume of fluid (VOF) methods \cite{rudman1997volume,Rider1997Reconstructing,Gueyffier1999Volume,Scardovelli2000Analytical,Pilliod2004Second} may be the most popular ones. The VOF methods use the volume of fluid function, which represents the fraction of the liquid volume in a given computational cell, to capture the interface position. The VOF function is updated by solving the advection equation with the finite volume scheme so that the mass conservation is ensured. Nevertheless, there are some problems in computing the geometrical variables such as the normal vector and curvature because the VOF function is not continuous. 

The other approach is to track the interface with a non-smearing representation, i.e., localize the interface as well as the interface interactions at an infinitely thin region. In the arbitrary Lagrangian Eulerian(ALE) methods \cite{Hirt1974arbitrary,Ling2010numerical}, the interface is represented by the computational mesh which evolves and deforms with the flow while it is represented by interface markers in the front-tracking method \cite{unverdi1992front}. Introduced by Osher et al. \cite{osher1988fronts}, the level-set methods share algorithmic similarities with the diffuse-interface methods and are simple to be implemented. Different from those sharp-interface mehtods mentiond above, the level-set methods employ a signed distance function known as the level-set function to represent the interface implicitly and are able to handle interface deformations and topological changes in a straightforward way \cite{Sussman1994level,Osher2001Level}. The main drawback of the level-set methods is the lack of discrete conservation properties. To address this issue, Hu et al. \cite{hu2006conservative} introduce a fully conservative level-set based sharp-interface method for compressible flows. The standard finite volume scheme on Cartesian grids is used for the far interface region while it is slightly modified for the near interface region. Unlike those hybrid methods such as the particle level-set methods \cite{Enright2002Hybrid,Wang2009improved} and the coupled level-set and volume of fluid methods(CLSVOF) \cite{M2007Coupling,Sussman2000Coupled}, this method maintains the simplicity of the pure level-set method and can be simply implemented in multi-dimension and multi-level time integrations. In the method of Hu et al., the treatments of surface tension and the viscous force at the interface singularity are not considered, which often play important roles in mulitphase flows. It is a challenge for numerical models of multiphase flows to handle the singular surface tension force only acting at the interface and the jump of material properties across the interface such as density and viscosity. By employing a weakly compressible model, Luo et al. \cite{luo2015conservative} propose a conservative sharp-interface method based on the work of Hu et al. for incompressible multiphase flows. In the cell cut by the interface, an effective viscosity is constructed to model the viscous flux across the interface and a constrained Riemann problem \cite{bo2011robust} is solved to model the interfacial flux caused by surface tension.

When explicit time discretization is adopted, the simulation suffers from the limitation of time step due to numerical stability reasons, i.e., CFL conditions such as $\Delta t  \propto \frac{1}{|\bm{u}|+c}$ for advection terms and $\Delta t  \propto \frac{\rho \Delta x^2}{\mu}$ for diffusive terms. In many multiphase flows problems, the time-scale features of different fluids are very different because of the differences of density, sound speed, viscosity, etc. A bottleneck of traditional multiphase simulation methods is that the fluid with small time-scale features imposes a small time step which is also used to evovle the states of fluid with larger time-scale features. To alleviate the stringent CFL restriction, the semi-implicit methods \cite{Kwatra2009Method,Kadioglu2008Adaptive,Yabe2007Unified} are proposed which are suitable for flows containing low Mach number regions. The calculation is divided into two parts: advection and non-advection. Since the non-advection terms containing the acoustic components are solved implicitly, the time step is constrained by material velocity only while the stringent CFL time step restriction imposed by the acoustic waves is avoided. However, in addition to leading to extra oscillations, these semi-implicit methods require lots of changes to the original code when applied to those multiphase flows solvers based on explicit time discretization. When the interface between the fluids can be considered as a free-boundary, the computation can be limited to the more viscous and dense fluid. For example, in \cite{Carrica2010Unsteady, Kim2003FREESURFACE}, the computation takes place only in the water phase and the computational time for the air phase is saved. Nevertheless, this method is not suitable for problems in which the air phase gets pressurized or the stresses on the liquid caused by the air phase cannot be ignored. 
  
The objective of this paper is to develop an accelerated sharp interface method based on \cite{hu2006conservative,luo2015conservative} for compressible and weakly compressible multiphase flows. Without generating unphysical oscillations, this method is relatively convenient to be coded up based on the original program. In contrast to the free-surface conditions, the present method makes no assumptions for the gas phase (or the less viscous and dense phase). The main idea of the present method is to obtain speedup by introducing larger time step for the evolution of fluid with larger time-scale features without violating the numerical stability requirements. The intermediate states of fluid with larger time-scale features are obtained by interpolation for the interaction step between two fluids, which is solved for synchronization. In addition, an interfacial flux correction is implemented to maintain the conservative property. The present method can be combined with a wavelet-based adaptive multi-resolution algorithm \cite{han2014adaptive} to achieve additional computational efficiency. For those multiphase flows problems that fluid with lager time-scale features occupies most of the computational domain, significant speedup can be obtained by using the present method because the fluid states of most computational cells will be updated with a larger time step. 

The paper is organized as follows: In section \ref{sec2}, we briefly review the original conservative sharp interface methods for compressible and weakly compressible multiphase flows. Subsequently we introduce that how to implement this accelerated method on a uniform grid as well as a multi-resolution grid in section \ref{sec3}. To validate the accelerated method, a number of numerical tests are carried out in section \ref{sec4}. Finally, the concluding remarks are given in section \ref{sec5}.

\section{Conservative sharp interface method}
\label{sec2}

%
%%%%%%%%%%%%%%%%%%%%%%%%%%%%%%%%%%%%%%%%%%%%%%%%%%%%%%%%
\begin{figure}[p]
\centering
\includegraphics[width=0.6\textwidth]{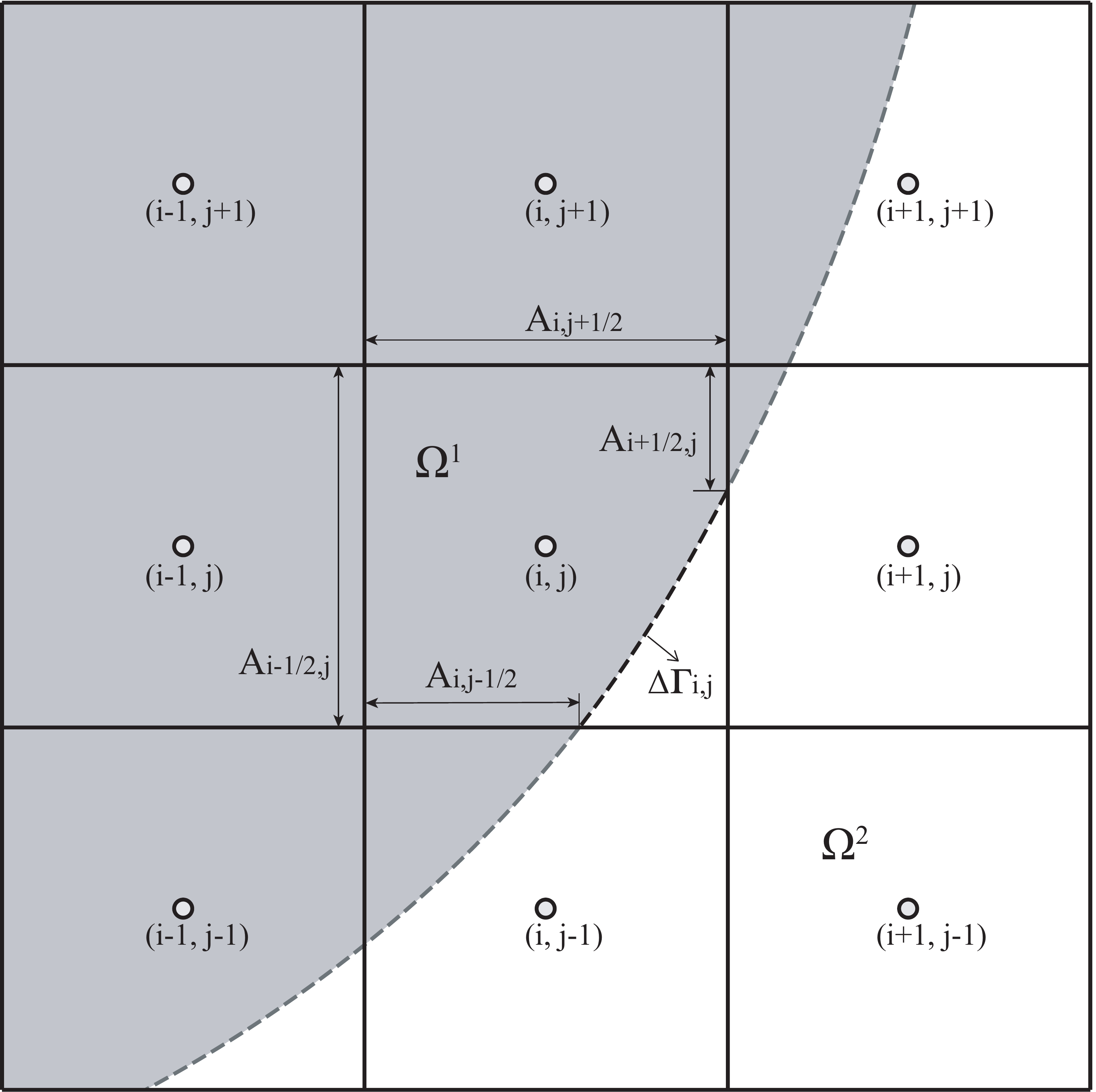}
\caption{Two dimensional schematic of conservative discretization for a cut cell $(i, j)$. The gray domain $\Omega ^ 1$ and the white domain $\Omega ^ 2  $ are occupied by fluid1 and fluid2, respectively.}
\label{Fig:geometric}
\end{figure}
%%%%%%%%%%%%%%%%%%%%%%%%%%%%%%%%%%%%%%%%%%%%%%%%%%%%%%%%
%

%compressible part
\subsection{Compressible sharp interface method}
\label{subsec2.1}
For the inviscid and compressible fluid, the governing equations can be written as
\begin{equation}
\frac{\partial \bm{U}}{\partial t}+\nabla \cdot \bm{F} = 0 ,
\label{eq:EulerEquation}
\end{equation}
where $ \bm{U} = {(\rho, \rho \bm{v}, E)}^{T}$ is the vector of mass, momentum and  total energy densities with the relation $E=\rho e+\frac{1}{2}\rho{|\bm{v}|}^{2}$, and $\bm{F}$ denotes the corresponding physical flux functions of $\bm{U}$. This set of equations is closed by an equation of state (EOS), which gives the thermodynamic properties of the materials. As shown in Fig. \ref{Fig:geometric}, the flow domain $\Omega$ is separated by an interface $\Gamma  (t)$ into two sub-domains ${\Omega}^{1}$ and ${\Omega}^{2}$, which represent the regions occupied by fluid1 and fluid2, respectively. As in Hu et al. \cite{hu2006conservative}, with explicit first-order forward time difference, the discretization of Eq. (\ref{eq:EulerEquation}) for each phase on a two-dimensional Cartesian grid with spacing $\Delta x$ and $\Delta y$ can be written as
\begin{equation}
\begin{aligned}
{\alpha}^{n+1}_{i,j}{\bm{U}}^{n+1}_{i,j}=&{\alpha}^{n}_{i,j}{\bm{U}}^{n}_{i,j}+\frac{\Delta t}{\Delta x \Delta y}\hat{\bm{X}}(\Delta {\Gamma}_{i,j})+\frac{\Delta t}{\Delta x}[{A}_{i-1/2,j}{\hat{\bm{F}}}_{i-1/2,j}-{A}_{i+1/2,j}{\hat{\bm{F}}}_{i+1/2,j}] \\
&+\frac{\Delta t}{\Delta y}[{A}_{i, j-1/2}{\hat{\bm{F}}}_{i, j-1/2}-{A}_{i, j+1/2}{\hat{\bm{F}}}_{i, j+1/2}].
\end{aligned}
\label{eq:Eulerdiscrete}
\end{equation}

Here, $\Delta t$ is the time step size, and ${\alpha}_{i,j}{\bm{U}}_{i,j}$ are the conservative variables in the cut cell, where ${\alpha}_{i,j}$ is volume fraction and ${\bm{U}}_{i,j}$ denotes cell-averaged densities of mass, momentum and energy of the considered fluid. ${A}_{i-1/2,j}$, ${A}_{i+1/2,j}$, ${A}_{i, j-1/2}$ and ${A}_{i, j+1/2}$ are cell-face apertures as shown in Fig. \ref{Fig:geometric}. $\hat{\bm{F}}$ denotes the reconstructed cell-face numerical flux obtained by a high-order shock-capturing scheme such as the WENO \cite{shu1988efficient} scheme. $\hat{\bm{X}}(\Delta {\Gamma}_{i,j})$ , where $\Delta {\Gamma}_{i,j}$ is the interface segment within the cut cell, denotes the inviscid momentum and energy flux across the interface determined by the interface interaction.

%weakly compressible part
\subsection{Weakly compressible sharp interface method}
\label{subsec2.2}
For the weakly compressible model \cite{luo2015conservative}, the governing equations can be written as
\begin{equation}
\frac{\partial \bm{U}}{\partial t}+\nabla \cdot \bm{F} =\nabla \cdot \bm{F}_{v}+\bm{a}_{s} ,
\label{eq:NSweaklycompressible}
\end{equation}
where the right-hand side terms $\bm{F}_{v}$ and $\bm{a}_{s}$, which don't exist in Eq. (\ref{eq:EulerEquation}), are viscous fluxes and  fluxes caused by surface tension, respectively. Note that the right-hand side terms in Eq. (\ref{eq:NSweaklycompressible}) will vanish when it represents the continuity equation. As shown in Luo et al. \cite{luo2015conservative}, Tait's equation is used as the artificial equation of state to close the governing equations:
\begin{equation}
\begin{aligned}
p&=B[(\frac{\rho}{\rho_{0}})^{\gamma}-1]+p_{0}&\\
B&=\frac{\rho_{0}{|\bm{v}_{f}|}^{2}}{\gamma M^{2}}&
\end{aligned}
\label{Eq: Tait's equation}
\end{equation}

Here, $\gamma$ is the the artificial specific-heat ratio and $\bm{v}_{f}$ represents the characteristic flow velocity. Moreover, $\rho_{0}$ and $p_{0}$ are the reference density and  pressure, respectively. The artificial sound speed is $c = \sqrt{\frac{\gamma B}{\rho}}$ and $M = \frac{|\bm{v}_{f}|}{c}$ denotes the Mach number. To impose incompressibility, the Mach number should be small enough, which handled by choosing the parameters $\gamma$, $B$ and $\rho_{0}$ specifically. As in section \ref{subsec2.1}, applying explicit first-order forward time difference to Eq. (\ref{eq:NSweaklycompressible}) leads to the following form:
\begin{equation}
\begin{aligned}
{\alpha}^{n+1}_{i,j}{\bm{U}}^{n+1}_{i,j}=&{\alpha}^{n}_{i,j}{\bm{U}}^{n}_{i,j}+\frac{\Delta t}{\Delta x \Delta y}[\hat{\bm{X}}(\Delta {\Gamma}_{i,j})+\hat{\bm{X}_{v}}(\Delta {\Gamma}_{i,j})+\hat{\bm{X}_{s}}(\Delta {\Gamma}_{i,j})]\\
&+\frac{\Delta t}{\Delta x}[{A}_{i-1/2,j}{\hat{\bm{F}}}_{i-1/2,j}-{A}_{i+1/2,j}{\hat{\bm{F}}}_{i+1/2,j}] \\
&+\frac{\Delta t}{\Delta y}[{A}_{i, j-1/2}{\hat{\bm{F}}}_{i, j-1/2}-{A}_{i, j+1/2}{\hat{\bm{F}}}_{i, j+1/2}]\\
&+\frac{\Delta t}{\Delta x}[{A}_{i-1/2,j}{\hat{\bm{F}_{v}}}_{i-1/2,j}-{A}_{i+1/2,j}{\hat{\bm{F}_{v}}}_{i+1/2,j}] \\
&+\frac{\Delta t}{\Delta y}[{A}_{i, j-1/2}{\hat{\bm{F}_{v}}}_{i, j-1/2}-{A}_{i, j+1/2}{\hat{\bm{F}_{v}}}_{i, j+1/2}].
\end{aligned}
\label{eq:NSdiscrete}
\end{equation}

The meanings of the terms which are the same as those in Eq. (\ref{eq:Eulerdiscrete}) are not repeated here. In the above equation, $\hat{\bm{F}_{v}}$ denotes the reconstructed cell-face numerical viscous flux obtained by central difference, while 
$\hat{\bm{X}_{v}}$ and $\hat{\bm{X}_{s}}$ are the fluxes across the interface due to the viscous force and surface tension, respectively. Note that all the interfacial fluxes $(\hat{\bm{X}},\hat{\bm{X}_{v}},\hat{\bm{X}_{s}})$ in Eq. (\ref{eq:NSdiscrete}) will vanish when it represents the continuity equation. 

In this paper, to maintain numerical stability, we follow Hu et al. \cite{hu2006conservative} and mix the conservative variables of small cut cells with their corresponding neighbors of larger volume fraction. The interface interaction models and more details about the conservative sharp interface methods can be found in \cite{hu2004interface,hu2006conservative,luo2015conservative}.

%level set part
\subsection{Level-set method}
\label{subsec2.3}
The level-set funtion $\Phi(\bm{x},t)$ \cite{osher1988fronts,sussman1998improved} is used to calculate the geometrical variables such as the volume fraction, cell-face apertures, the normal vector, etc. With $|\nabla \Phi = 1|$, the level-set function describes the signed distance from the interface to each cell center. The location of the interface is given by the zero level-set $\Phi = 0$ implicitly and the level-set field is updated by the linear advection equation \cite{fedkiw1999non}
\begin{equation}
\frac{\partial \Phi}{\partial t}+\bm{u} \cdot \nabla \Phi = 0,
\label{LS:linearadvection}
\end{equation}
where $\bm{u}$ denotes the level-set advection velocity obtained by the interface interaction model \cite{hu2004interface}.

In practice, a reinitialization procedure is needed for maintaining the signed distance property of the level-set function, which will be violated during the evolution of the interface. Besides, in order to complete the interpolation stencils for the near interface cells and solve the interface interaction, we need an extending algorithm to extrapolate the fluid states.
 
%the acceleration algortihm
\section{The accelerated method}
\label{sec3}
Considering a flow domain occupied by fluid1 and fluid2, we can obtain time step of fluid1 ${\Delta t}_{1}$ and time step of fluid2 ${\Delta t}_{2}$ according to CFL conditions. With the assumption that ${\Delta t}_{1}$ is much larger than ${\Delta t}_{2}$, the ratio of time steps is defined as $n = {\Delta t}_{1}/{\Delta t}_{2}$. Note that the second-order TVD Runge-Kutta (RK2) scheme \cite{shu1988efficient} is used as the time integration method in this paper.
\subsection{Uniform grid}
\label{subsec3.1}
%
%%%%%%%%%%%%%%%%%%%%%%%%%%%%%%%%%%%%%%%%%%%%%%%%%%%
\begin{figure}[p]
\centering
\includegraphics[width=0.6\textwidth]{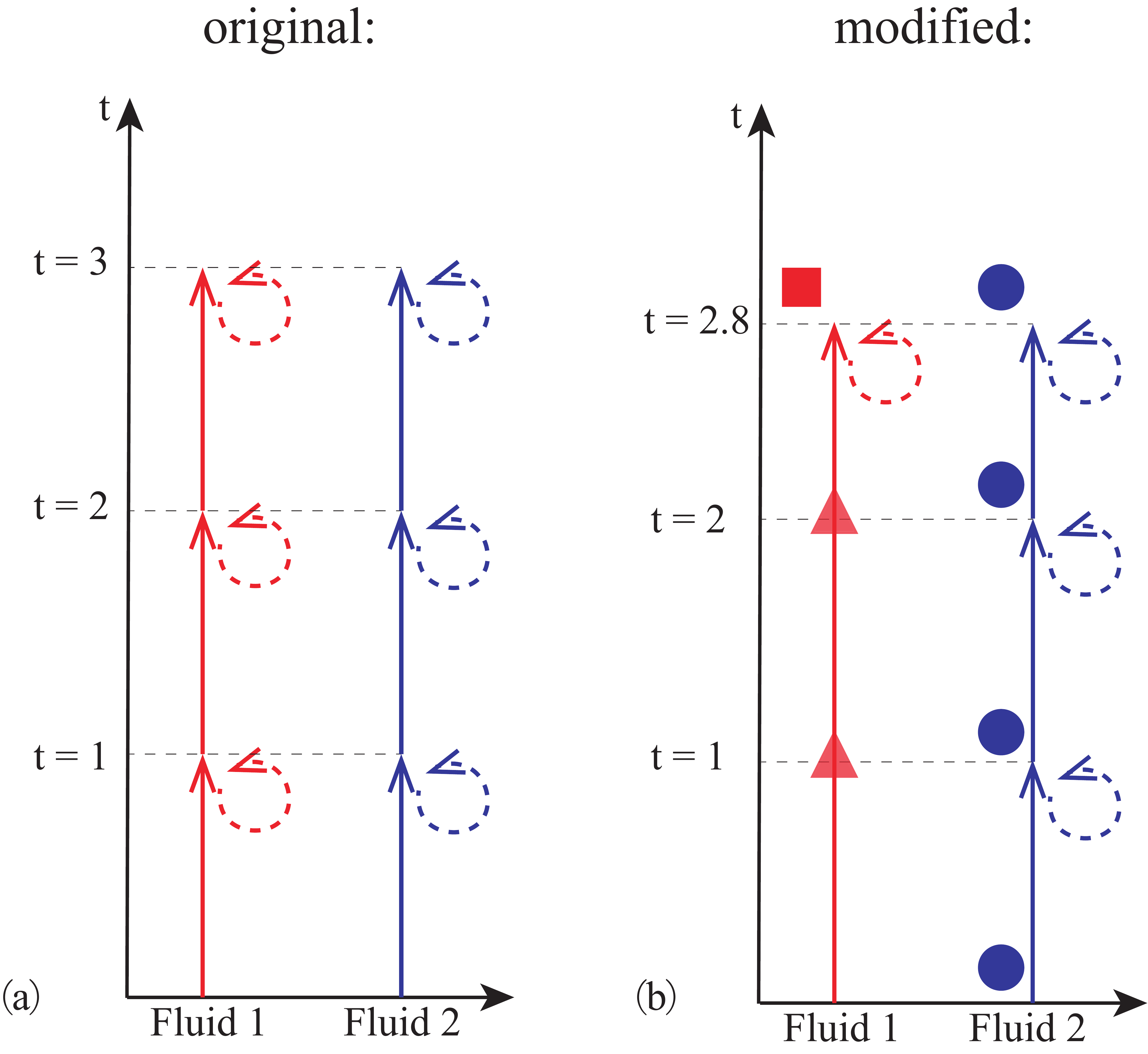}
\caption{Schematic of the accelerated method on a uniform grid for one time cycle: (a) The original method; (b) The accelerated method. Solid arrows and dashed arrows denote the first step and the second step of RK2, respectively. In the original method, the states of both fluid1 and fluid2 are updated with the time step ${\Delta t}_2 = 1$ while in the accelerated method ${\Delta t}_{1} = 2.8$ and ${\Delta t}_2 = 1$ are used for the evolution of fluid1 and fluid2, respectively. The intermediate states of fluid1 are obtained by interpolation (represented by filled triangles). To maintain the conservative property, the interfacial fluxes transferred to fluid2 are recorded (represented by filled circles) during the whole time cycle and an interfacial flux correction is made for fluid1 (represented by filled squares) at last.}
\label{Fig:algorithm}
\end{figure}
%%%%%%%%%%%%%%%%%%%%%%%%%%%%%%%%%%%%%%%%%%%%%%%%%%%
%
When the accelerated method is implemented on a uniform grid, it does not matter whether $n$ is an integer or not. However, when $n$ is a decimal number, the computational efficiency will be influenced if the decimal part of $n$ is not large enough. To address this issue, $n$ is truncated to an integer by reducing ${\Delta t}_{1}$ when the decimal part of $n$ is less than the preset threshold. In this paper, we choose 0.7 as the threshold for all cases. In one time cycle of the accelerated method, value of ${\Delta t}_{1}$ will be kept while value of ${\Delta t}_{2}$ will be updated along with the calculation. We assume that ${\Delta t}_{2}$ is always equal to the same value for simplicity. With the assumption that ${\Delta t}_{1}=2.8$ and ${\Delta t}_{2}=1$, the basic ideas of the accelerated method for one time cycle are illustrated in Fig. \ref{Fig:algorithm}. Note that some procedures such as the extending step, the interaction step, the mixing step, which can be found in \cite{hu2006conservative,luo2015conservative}, and the evolution of the level-set field are not drawn in this figure. 

First, after solving the interaction step for obtaining interfacial fluxes, we compute the first step of RK2 for fluid1 with ${\Delta t}_{1}=2.8$. Both previous and future conservative variables of fluid1 are stored for interpolating its intermediate states. Second, we compute the first step of RK2 for fluid2 and the level-set field with the time step ${\Delta t}_{2}=1$. Let $t_1$ and $t_2$ denote the evolution time of fluid1 and fluid2, respectively. In one time cycle, we have $t_1 = {\Delta t}_{1}$ while $t_2$ is accumulated every time ${\Delta t}_{2}$ is updated according to CFL condition. The intermediate states of fluid1 corresponding to fluid2 can be obtained by 
\begin{equation}
(\alpha \bm{U})_{1}^{interpolation} = (1-{\frac{t_{2}}{t_{1}}}) \, (\alpha \bm{U})_{1}^{previous}+{\frac{{t_{2}}}{t_{1}}} \, (\alpha \bm{U})_{1}^{future},
\label{Eq:interpolation}
\end{equation} 
Then, with flow variables of fluid1 at $t_1 = 1$ obtained by interpolation, the interaction step is solved so that we can compute the second step of RK2 for fluid2 and the level-set field. The interfacial fluxes transferred to fluid1 and fluid2 are not equal due to the separated evolution of two fluids. To maintain the conservative property, we record the interfacial fluxes transferred to fluid2 during the whole time cycle and make the interfacial flux correction for fluid1 at last. The correction quantities for both inviscid and viscous interfacial fluxes can be calculated according to two different situations as follows:

i. if the interfacial fluxes will be used to update the states of both fluid1 and fluid2
\begin{equation}
\begin{aligned}
\hat{\bm{X}}^{correction} &= (\hat{\bm{X}}^{correction})^{*}-0.5 \times\hat{\bm{X}}\times {\Delta t}_{1} + 0.5 \times\hat{\bm{X}} \times {\Delta t}_{2} 
\\
\hat{\bm{X}_{v}}^{correction} &= (\hat{\bm{X}_{v}}^{correction})^{*}-0.5 \times\hat{\bm{X}_{v}}\times {\Delta t}_{1} + 0.5 \times\hat{\bm{X}_{v}} \times {\Delta t}_{2} 
\end{aligned}
\label{Eq:firstcorrection}
\end{equation}

ii. if the interfacial fluxes will only be used to update the states of fluid2
\begin{equation}
\begin{aligned}
\hat{\bm{X}}^{correction} &= (\hat{\bm{X}}^{correction})^{*}+ 0.5 \times\hat{\bm{X}} \times {\Delta t}_{2} 
\\
\hat{\bm{X}_{v}}^{correction} &= (\hat{\bm{X}_{v}}^{correction})^{*}+ 0.5 \times\hat{\bm{X}_{v}} \times {\Delta t}_{2} 
\end{aligned}
\label{Eq:secondcorrection}
\end{equation}
Note that the interfacial flux correction quantities should be set to 0 at the beginning of one time cycle and the superscript $*$ denotes the previous values. The calculation of the correction quantities for interfacial fluxes due to surface tension is slightly different. Caused by mechanical equilibrium, surface tension will result in a pressure jump at the interface, which can be expressed by
\begin{equation}
[\,p\,]_{\Gamma}=p_{I,1}-p_{I,2} = \sigma\kappa,
\label{Eq:surfacetension}
\end{equation}
where $\sigma$, $\kappa$ are the surface tension coefficient and curvature, respectively. As in Luo et al. \cite{luo2015conservative}, the interfacial fluxes due to surface tension $\hat{\bm{X}_{s}}$ are integrated into the inviscid interfacial fluxes $\hat{\bm{X}}$, and then we have $\hat{\bm{X}_{1}}$ for fluid1 and $\hat{\bm{X}_{2}}$ for fluid2 with the relation $\frac{\hat{\bm{X}_{1}}}{\hat{\bm{X}_{2}}} =  \frac{p_{I,1}}{ p_{I,2}} $. The correction quantities for interfacial fluxes due to surface tension can be obtained by:

i. if the interfacial fluxes will be used to update the states of both fluid1 and fluid2
\begin{equation}
\hat{\bm{X}_{s}}^{correction} = (\hat{\bm{X}_{s}}^{correction})^{*}-0.5 \times\hat{\bm{X}_{1}}\times {\Delta t}_{1} + 0.5 \times\frac{p_{I,1}}{ p_{I,2}}\times \hat{\bm{X}_{2}} \times {\Delta t}_{2} 
\label{Eq:firstcorrection_surfacetension}
\end{equation}

ii. if the interfacial fluxes will only be used to update the states of fluid2
\begin{equation}
\hat{\bm{X}_{s}}^{correction} = (\hat{\bm{X}_{s}}^{correction})^{*} + 0.5 \times\frac{p_{I,1}}{ p_{I,2}}\times \hat{\bm{X}_{2}} \times {\Delta t}_{2} 
\label{Eq:secondcorrection_surface_tension}
\end{equation}

Next, we repeat updating the states of fluid2 and the level-set field until $t_{2}$ is greater than (or equal to) $t_1$. At the end of one time cycle, $t_2$ should be equal to $t_1$ for synchronization. Hence, we truncate ${\Delta t}_2$ by ${\Delta t}_{2}=t_1-{t_2}^{*}$ and compute the first step of RK2 for fluid2 and the level-set field with it. Finally, after computing the second step of RK2 for fluid1, fluid2 and the level-set field together, we make the interfacial flux correction for fluid1.

In the original method, the minimum time step i.e. ${\Delta t}_{2}=1$ is used to evolve the fluid states of both fluid1 and fluid2. In order to advance the flow field to $t = 3$, we have to compute RK2 three times for both fluid1 and fluid2. With the accelerated method, we compute RK2 three times for fluid2 and only once for fluid1 to advance the flow field to $t=2.8$. Compared with the RK2 scheme, time costs of new extra procedures such as interpolation, interfacial fluxes recording, interfacial flux correction are much lower and can almost be omitted. In the original method, let $m$ denotes the proportion of the calculation time of RK2 for fluid1 in the total calculation time of marching a time step. The speedup ratio $r_s$, which is defined as the ratio of previous total simulation time to modified total simulation time, can be calculated by
\begin{equation}
r_s=\frac{n}{[n]-([n]-1)m},
\label{Eq:speedup_ratio}  
\end{equation}
where $[\ ]$ denotes rounding a number to the next larger integer.

With the assumption that ${\Delta t}_1 > {\Delta t}_2$ , the procedures of the accelerated method for a whole time cycle can be summarized as follows:
\begin{enumerate}
\item Extend the fluid states and calculate interfacial fluxes with the interface interaction models proposed in \cite{hu2006conservative,luo2015conservative}. Then obtain intefacial flux correction quantities via Eq. (\ref{Eq:firstcorrection}) or Eq. (\ref{Eq:firstcorrection_surfacetension}) according to whether the surface tension is considered.
\item Compute the first step of RK2 for fluid1 with ${\Delta t}_1$ and store both previous and future conservative variables of fluid1 for interpolating its intermediate states via Eq. (\ref{Eq:interpolation}).
\item Compute the first step of RK2 for fluid2 and the level-set field with ${\Delta t}_2$. With the interpolated intermediate states of fluid1, extend fluid states, solve the interaction step and update interfacial flux correction quantities via Eq. (\ref{Eq:secondcorrection}) or Eq. (\ref{Eq:secondcorrection_surface_tension}) according to whether the surface tension is considered. 
\item Compute the second step of RK2 for fluid2 and the level-set field. Then update ${\Delta t}_2$ according to CFL conditions while ${\Delta t}_1$ is kept, and accumulate $t_2$.
\item Repeat steps 3-4 until $t_2$ is greater than (or equal to) $t_1$. Truncate ${\Delta t}_2$ and implement the step 3. Note that the interfacial flux correction quantities should be updated via Eq. (\ref{Eq:firstcorrection}) or Eq. (\ref{Eq:firstcorrection_surfacetension}).  Compute the second step of RK2 for fluid1, fluid2 and the level-set field together and finally make the interfacial flux correction for fluid1.
\end{enumerate}

\subsection{Multi-resolution grid}
\label{subsec3.2}
The original MR algorithm \cite{han2014adaptive} employs a storage-and-operation-splitting pyramid data structure which will be adaptively updated through the multi-resolution analysis. The pyramid data structure is defined as a set of tables in which the basic element is called a node. There are two types of nodes, empty node which means there is no computational data and occupied node which contains a block of $n\times m$ computational cells. When it has no substructures, a node is called a leaf, which can be viewd as a sub-domain with a uniform grid and the simulations are carried out on. First, by using a projection operator $P_{l+1 \rightarrow l}$ \cite{Domingues2008adaptive, Bihari1997multiresolution}, cell-averages at the level $l$ are  calculated from its child-level $l+1$ exactly. Then a prediction operator $P_{l \rightarrow l+1}$ \cite{Domingues2008adaptive, Bihari1997multiresolution} is employed to give an approximation of cell-averages at the level $l+1$ by using those values obtained by projection at the level $l$. The prediction errors between the approximation and the exact cell-averages at the level $l+1$ is called details. When the detail is smaller than a certain threshold, the flow filed can be represented by the values on the coarser grid, otherwise the corresponding nodes need to be refined. In order to improve the computational efficiency, the local time-stepping scheme \cite{Domingues2008adaptive} is implemented. The main principle is that the finer leaves are updated more frequently than the coarser leaves, which is handled by introducing larger time step for the evolution of the coarser leaves, e.g., the time step of leaves at the level $l$ is twice the time step of leaves at the level $l+1$. The intermediate states of the coarser leaves are obtained by interpolation. At the coarse-fine resolution boundaries, the discrete conservation is violated because the fluxes at the interface of the coarser leaves are computed by using the cell-averaged values of the adjacent nodes, which are obtained by projecting from the child-level. To ensure the discrete conservation, a flux correction step is applied to the coarser leaves located at the coarse-fine resolution boundaries.

When the accelerated method is combined with the MR algorithm, those procedures of multi-resolution analysis such as projection, prediction and refinement can be implemented without modification. Since all multiphase leaves are located at the finest level according to the MR algorithm \cite{han2014adaptive}, those procedures corresponding to the accelerated method such as interpolation for fluid advanced with a larger time step, interfacial flux recording and interfacial flux correction can be applied to the multiphase leaves in a straightforward way, as no resolution jumps are involved. The time steps of fluid1 and fluid2 at the level $l$ are represented by ${\Delta t}_{1}^l$ and ${\Delta t}_{2}^l$, respectively, and the assumption that ${\Delta t}_{1}^l$ is much larger than ${\Delta t}_{2}^l$ is still held. Note that the time step of each node is set at the beginning and will be kept during the whole time cycle. With the relations that ${\Delta t}_{1}^l = 2 {\Delta t}_{1}^{l+1}$ and ${\Delta t}_{2}^l = 2 {\Delta t}_{2}^{l+1}$, the ratio of time steps $n$ at every level equals to the same value. To combined with the local time-stepping scheme, $n$ will be truncated to an integer by reducing ${\Delta t}_{1}^l$ when it is a decimal number. At the level $l$, the states of fluid1 will be updated $2^l$ times while the states of fluid2 will be updated $n \times 2^l$ times. Note that the minimum level is $l = 0$.

\begin{figure}[p] 
\centering
\includegraphics[width=0.8\textwidth]{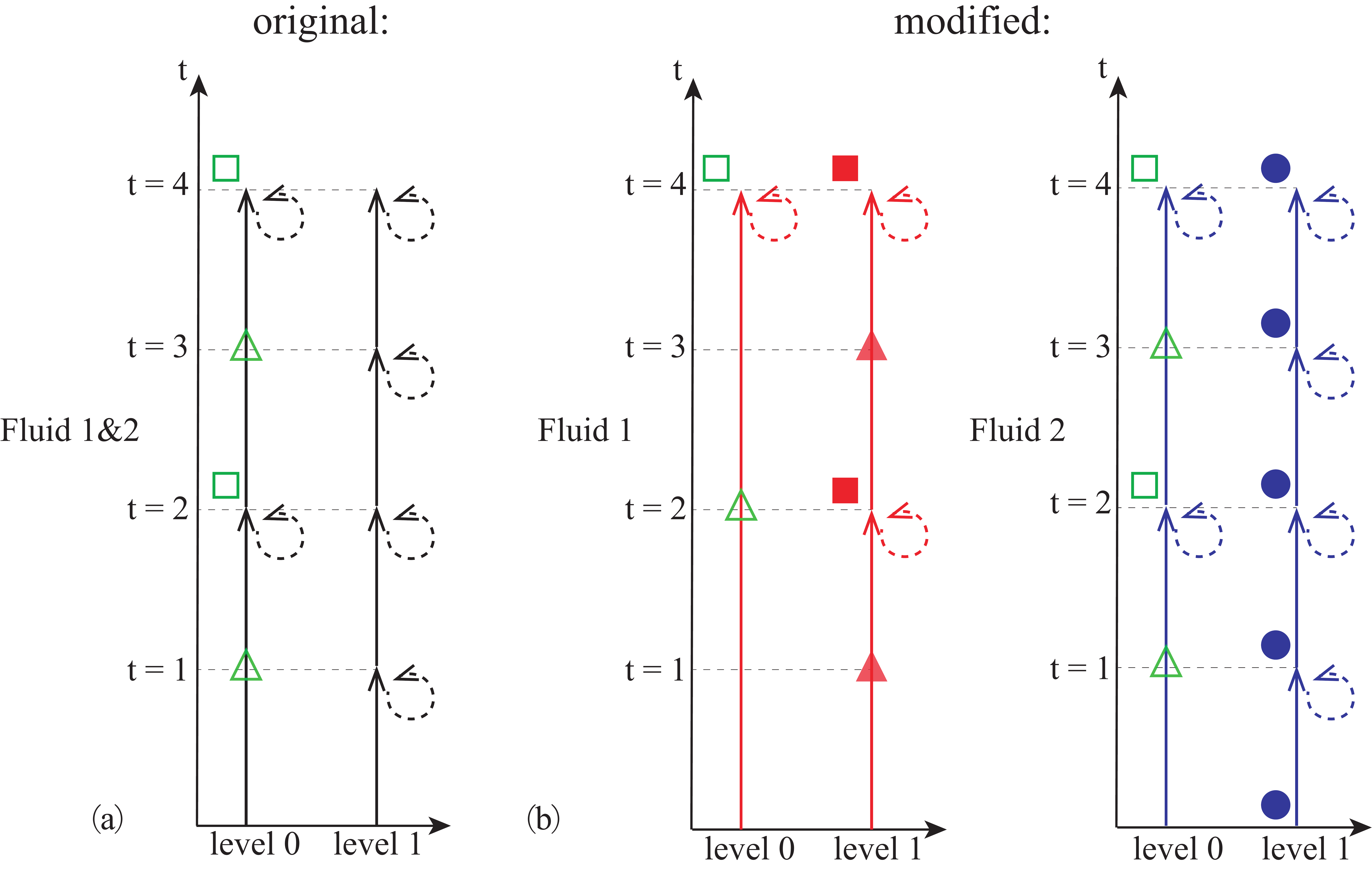}
\caption{Schematic of the accelerated method combined with the MR algorithm for one time cycle: (a) The original MR algorithm; (b) The accelerated method. The meanings of solid arrows, dashed arrows, filled squares, filled circles and filled triangles are the same as those in Fig. \ref{Fig:algorithm}, which represent the extra procedures related to the accelerated method. Since all multiphase leaves are located at the finest level, these procedures can be directly implemented at the finest level as same as on a uniform grid. 
The unfilled triangle and square denote the interpolation and the flux correction step, respectively, which are related to the local time-stepping scheme. }
\label{Fig:algorithm_multi}
\end{figure}

For simplicity, we only consider two scale levels (level 0 and level 1) here. With the assumption that ${\Delta t}_1^1 = 2$ and ${\Delta t}_2^1 = 1$, the basic ideas of the accelerated method combined with the MR algorithm for one time cycle are illustrated in Fig. \ref{Fig:algorithm_multi}. It can be observed that after being modified, the evolution of each fluid is almost the same as the original process except the extra procedures implemented at the finest level. As discussed in the above paragraph, these extra procedures related to the accelerated method are implemented at the finest level as same as on a uniform grid. Hence we do not discuss the repetitive details here. Let $L_{max}$ denotes the maximum level of the pyramid data structure and $I$ represents the counter of the cycle, which counts from zero. The general process of the accelerated method combined with the MR algorithm for one time cycle is given as follows:
\begin{enumerate}
\item Compute the first step of RK2 for leaves satisfying the conditions as follows:
$$
\left\{
\begin{aligned}
if\  &I \ \% \ 2^{L_{max}-l} \  = \  0 \quad \text{leaves of fluid1 at the level $l$}, \\
if\  &I\%(2^{L_{max}-l}\times n) \  = \  0 \quad \text{leaves of fluid2 at the level $l$}, \\
\end{aligned}
\right.
$$
where $\%$ denotes the modulus operator. 
\item Obtain intermediate states of both fluid1 and fluid2 at the coarser level and intermediate states of fluid1 at the finest level by interpolation. Record the corresponding fluxes for conservative correction.
\item Compute the second step of RK2 for leaves satisfying the conditions as follows:
$$
\left\{
\begin{aligned}
if\  &(I+1)\%2^{L_{max}-l} \  = \  0 \quad \text{leaves of fluid1 at the level $l$}, \\
if\  &(I+1)\%(2^{L_{max}-l}\times n) \  = \  0 \quad \text{leaves of fluid2 at the level $l$}. \\
\end{aligned}
\right.
$$
\item When two successive levels reach the same discrete time, a flux correction step is applied to those leaves located at the coarse-fine resolution boundaries. In addition, at the finest level, make the interfacial flux correction for fluid1 every time the evolution time of two fluids are the same. Then the counter $I$ is accumulated once.
\item Repeat steps $1-4$ until $I$ equals $2^{L_{max}}\times n $.
\end{enumerate}

%
%%%%%%%%%%%%%%%%%%%%%%%%%%%%%%%%%%%%%%%%%
%%%%%%%%%%%%%%%%%%%%%%%%%%%%%%%%%%%%%%%%%
%

\section{Numerical examples}
\label{sec4}
    The following numerical examples are provided to illustrate the potential of the accelerated sharp-interface method for compressible and weakly compressible multiphase flows. Different kinds of flow problems involving shock waves, the viscous force and surface tension  are calculated to compare the accuracy and speed of the present method with the original method. For all cases, with the CFL number of 0.6, one-phase calculations are carried out with the fifth-order WENO-LF method \cite{jiang1996efficient} and the second-order TVD Runge-Kutta scheme \cite{shu1988efficient}. In section \ref{subsec4.1} - section \ref{subsec4.2}, an ideal-gas EOS is used for gas while water is modelled with Tait's equation Eq. (\ref{Eq: Tait's equation}). In section \ref{subsec4.3} - section \ref{subsec4.7}, for the wealy compressible model, Tait's equation Eq. (\ref{Eq: Tait's equation}) is used as the artificial EOS for two phases.
    
\subsection{Gas-water interaction}
\label{subsec4.1}
Two 1D problems with the gas-water interaction are calculated to validate the present method for compressible multiphase flows. With respect to the state of water at 1 atmosphere and length scale 1 m, the non-dimensional parameters in Tait's equation are given by $B = 3310$, $\rho_0 = 1$ and $p_0 = 1$. In the two problems, simulations are carried out on a uniform mesh with 200 grid points and the reference solution is sampled on 1000 grid points. Due to the small calculation costs of the two problems, speedup is slight and will not be discussed here.

\subsubsection{Gas-water shock tube problem}
\label{subsubsec4.1.1}
\begin{figure}[p]
\centering
\includegraphics[width=1.0\textwidth]{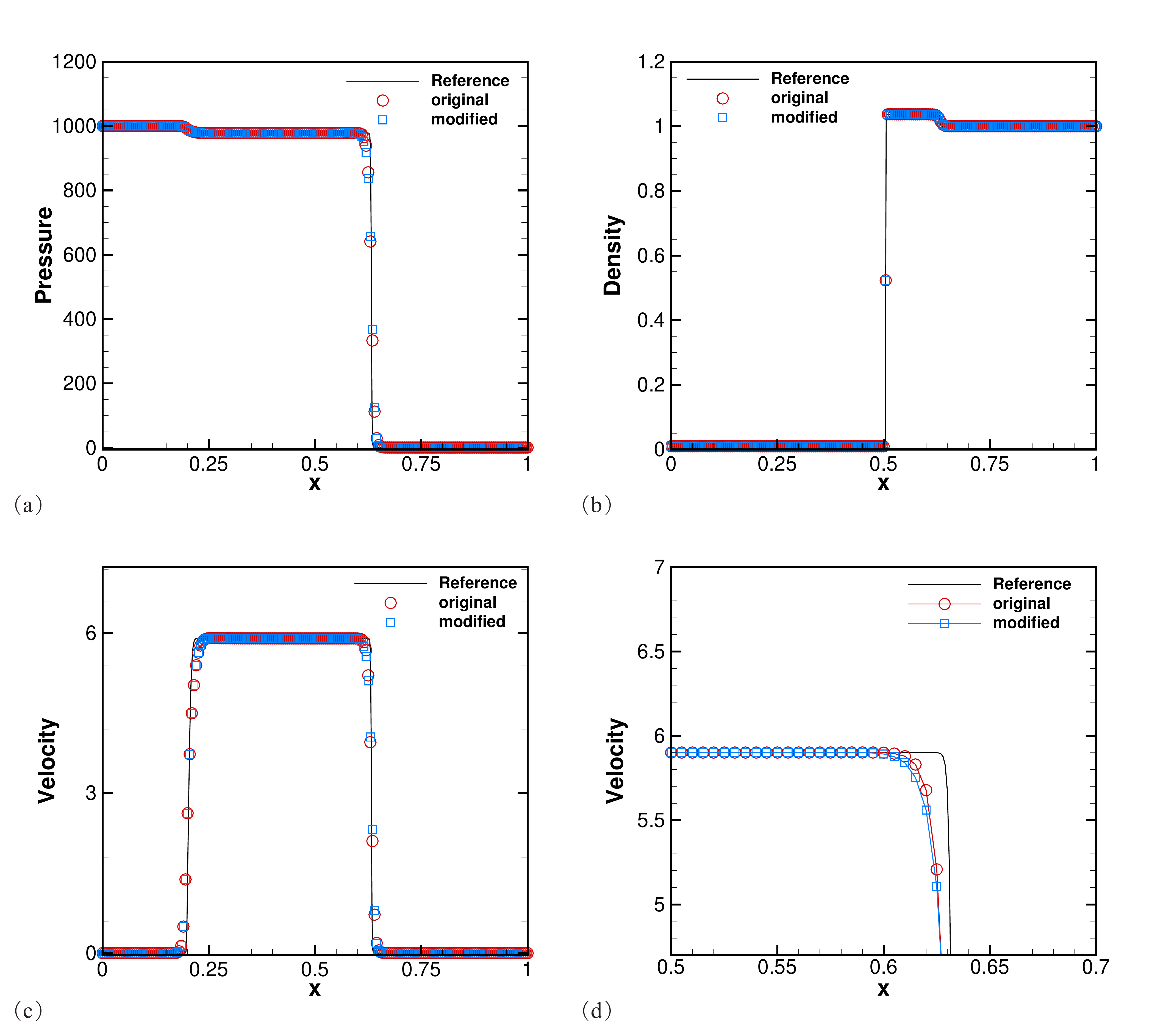}
\caption{Gas-water shock tube problem:  (a) Pressure profile;  (b) Density profile;  (c) Velocity profile;  (d) Enlarged velocity profile. The simulation is carried out on a uniform mesh with 200 grid points and the reference solution is sampled on 1000 grid points.  }
\label{Fig:gas-water shock tube}
\end{figure}

The initial conditions are given as
\begin{equation}
(\rho, u, p, \gamma)=\left\{
\begin{aligned}
&(0.01,0,1000,2&)\quad if\ x<0.5, \\
&(1,0,1,7.15&)\quad if\ x>0.5,
\end{aligned}
\right.
\end{equation}
and the final time is $t=0.0008$. In this problem, the transmitted and reflected wave fronts move rapidly while the high-pressure gas expands slowly. At first, the time step of water is about three times the time step of gas according to CFL conditions. Fig. \ref{Fig:gas-water shock tube} gives the computed pressure, velocity, density profiles of the present method, which show a good agreement with the original results. No oscillations are observed and the position of the gas-water interface is the same as that in the original result. The numerical dissipation increases because the states of water is evolved with a larger time step. As a result, the shock wave is underestimated slightly, which can be observed from the enlarged velocity profile Fig. \ref{Fig:gas-water shock tube}(d).
%
%%%%%%%%%%%%%%%%%%%%%%%%%%%%%%%%%%%%%%%%%%%%%%
%%%%%%%%%%%%%%%%%%%%%%%%%%%%%%%%%%%%%%%%%%%%%%
%

\subsubsection{ Shock impacting on an air-water interface}
\label{subsubsec4.1.2}
\begin{figure}[p]
\centering
\includegraphics[width=1.0\textwidth]{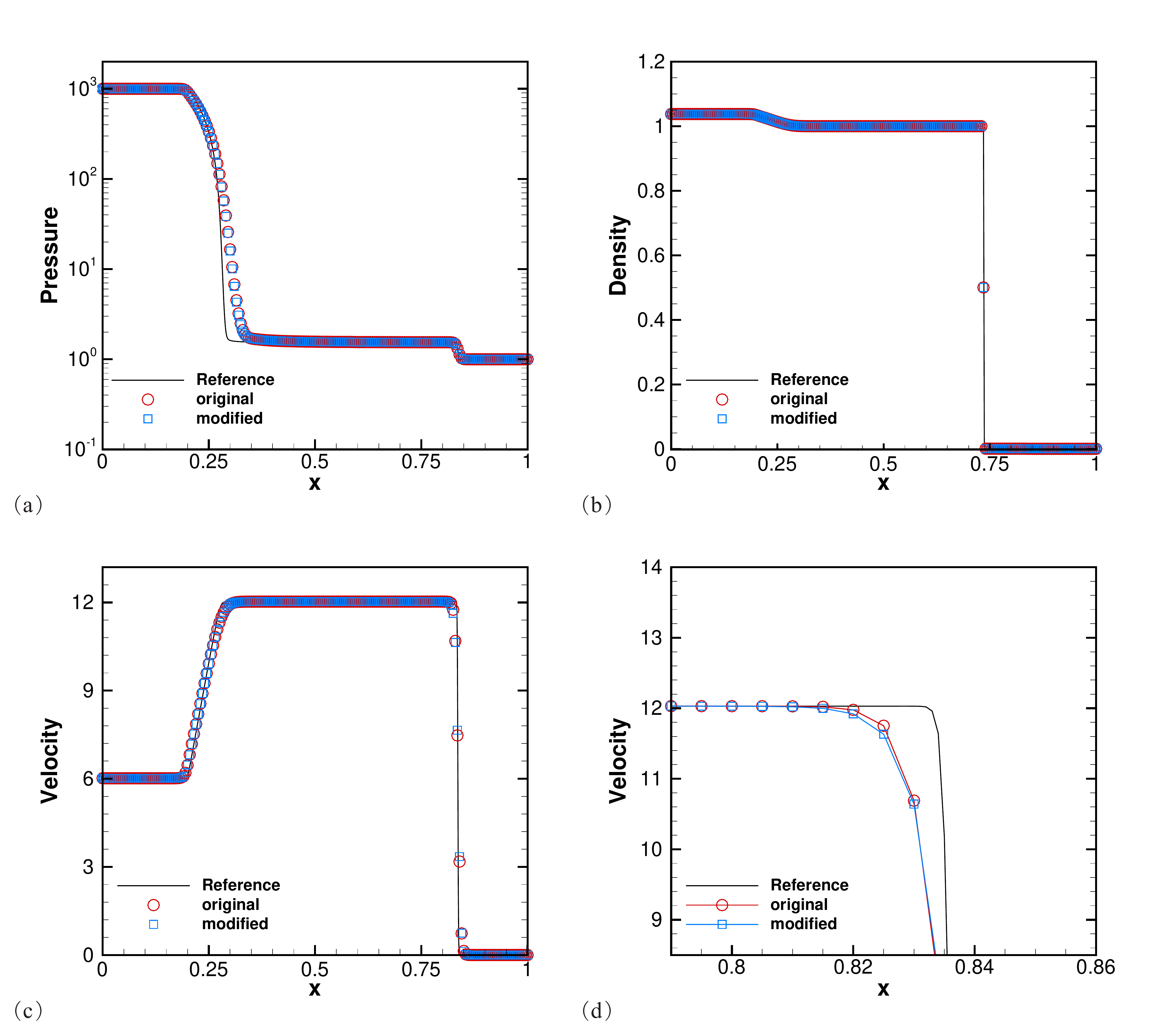}
\caption{Shock impacting on air-water interface:  (a) Pressure profile;  (b) Density profile;  (c) Velocity profile;  (d) Enlarged velocity profile. The simulation is carried out on a uniform mesh with 200 grid points and the reference solution is sampled on 1000 grid points.}
\label{Fig:air-bubble-collapse}
\end{figure}

Taken from \cite{liu2003ghost}, this case is an air bubble collapse problem in one dimension. The initial conditions are given as
\begin{equation}
(\rho, u, p, \gamma)=\left\{
\begin{aligned}
&(1.0376, 6.0151, 1000, 7.15&)\quad if\ 0<x<0.7, \\
&(0.001, 0, 1, 1.4&)\quad if\ 1>x>0.7, 
\end{aligned}
\right.
\end{equation}\\
and the final time is $t=0.003$. In this problem, the underwater shock wave impacts on the air-water interface so that a weak shock wave transmitted into the air and a very strong rarefaction wave reflected back to the water medium. At first, the time step of air is about four times the time step of water according to CFL conditions. In Fig. \ref{Fig:air-bubble-collapse}, the computed pressure, velocity, density profiles of the present method show a good agreement with the original results and no oscillations are observed. The air-water interface is captured at the position as same as in the original result. The shock wave is underestimated slightly because of larger numerical dissipation caused by introducing larger time step for air. 
%
%%%%%%%%%%%%%%%%%%%%%%%%%%%%%%%%%%%%%%%%%%%%%%
%%%%%%%%%%%%%%%%%%%%%%%%%%%%%%%%%%%%%%%%%%%%%%
%

\subsection{Periodic motion of a water droplet}
\label{subsec4.2}

We consider the translation of a water droplet in a periodic domain to study the convergence and the speedup of the present method. Embedded in the air, a water droplet of radius $ R=0.1\ m $ is placed at the center of a $ 1\ m\times 1\ m $ square computational domain with periodic boundary conditions. The reference length is $l_f = 1\ m$ and the reference velocity is $ v_{f}=\sqrt{p_{f} / \rho_{f}} $, where the reference pressure and density are given by $ p_{f}=1\ atm $, $\rho_{f}=1\ kg/m^{3} $, respectively. In this case, the non-dimensional parameters in Tait's equation are chosen as $B = 3310$, $\rho_0 = 1000$ and $p_0 = 1$. The initial conditions are given by
\begin{equation}
\left\{
\begin{aligned}
\rho&=1.2, p=1, u=\dfrac{\sqrt{10}}{100},v=0, \gamma=1.4&\quad air, \\
\rho&=1000, p=1, u=\dfrac{\sqrt{10}}{100},v=0, \gamma=7.15&\quad water, \\
\phi&=-0.1+\sqrt{(x-0.5)^{2}+(y-0.5)^{2}}&\quad level\ set, 
\end{aligned}
\right.
\end{equation}\\
and the final time is $t=10\sqrt{10}$ which is exactly a cycle. 

In Table. \ref{Table:watermovespeedup}, we list the speedup ratios for different resolutions. When the domain is discretized by a uniform mesh with increasing grid points from $ 32\times32 $ to $ 256\times256 $, it can be observed that the speedup ratio keeps increasing along with gird refinement. According to CFL conditions, the time step of gas is about four times the time step of water during the whole simulation in this case. As resolution growing, the ratio of the number of grid points occupied by the gas to the number of grid points occupied by the water increases so that $m$ defined in Eq. (\ref{Eq:speedup_ratio}) increases. Hence, according to Eq. (\ref{Eq:speedup_ratio}), the speedup ratio $r_s$ will increase along with grid refinement. With a uniform grid involving $128\times128\times128$ cells, the 3D version of this case is calculated by slightly modifying the initial conditions. The velocity component in the z-axis direction $w$ is set to 0 and the level-set field is initialized by $\Phi=-0.1+\sqrt{(x-0.5)^{2}+(y-0.5)^{2}+(z-0.5)^{2}}$. As shown in Table. \ref{Table:watermovespeedup}, with the resolution as same as in two dimensions, additional speedup ratio of about 0.5 is obtained in three dimensions. The reason why the speedup ratio increases is that the ratio of the number of grids points occupied by the gas to the number of grid points occupied by the water in three dimensions is larger than it in two dimensions when the resolutions are the same. 

\begin{figure}[p]
\centering
\includegraphics[width=1.0\textwidth]{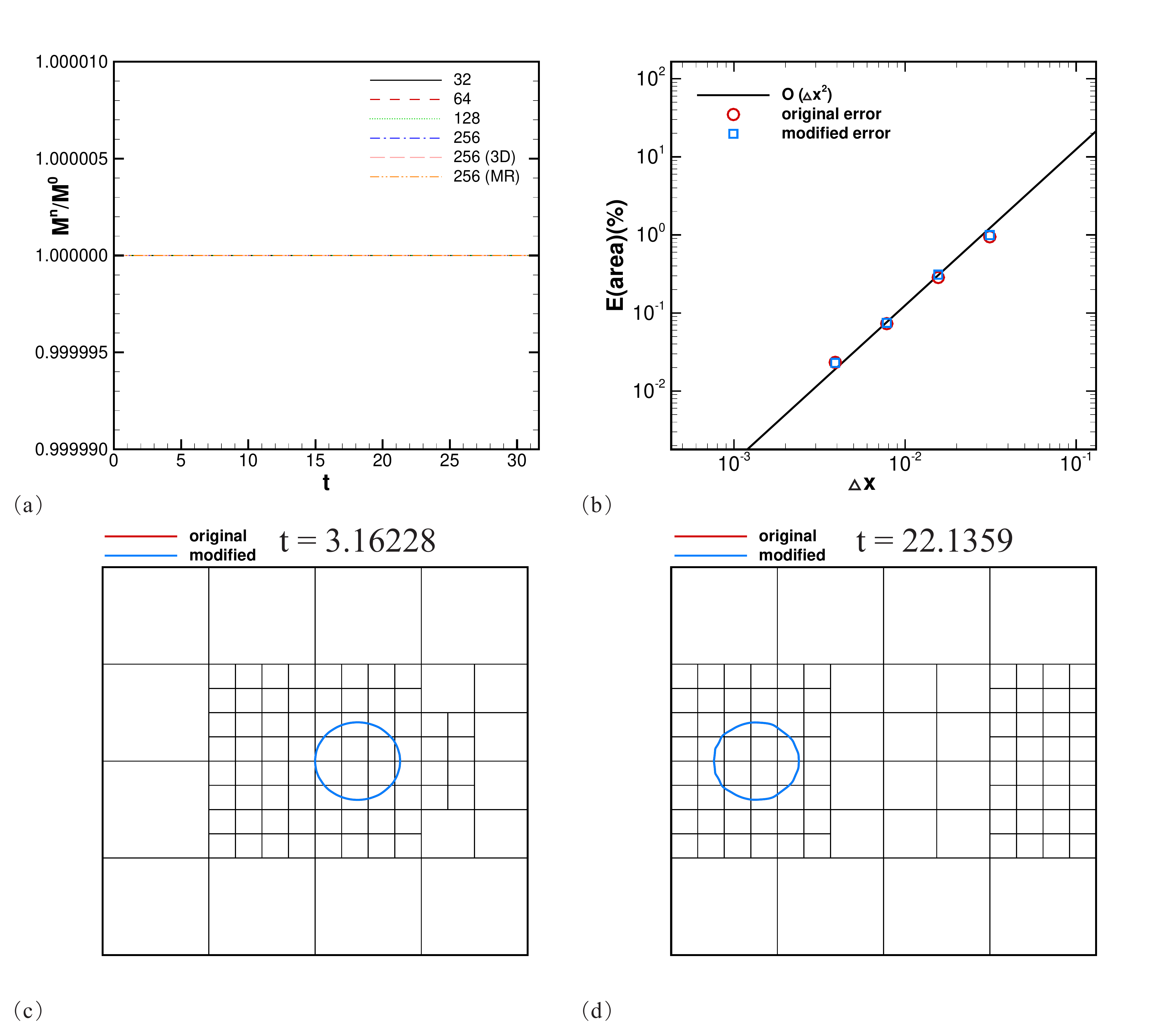}
\caption{Periodic motion of a water droplet:  (a) Time variation of droplet mass to its initial value;  (b) Relative error of droplet area according to resolution; (c) The multi-resolution representation at $t = 3.16228$; (d) The multi-resolution representation at $t = 22.1359$. (b) is obtained on the uniform mesh with grid refinement while (c) and (d) are obtained on the multi-resolution grid where the effective grid resolution at the finest level is $256 \times 256$.}
\label{Fig:water-periodic-movement}
\end{figure}

After one cycle, the water droplet returns the original position and the exact area of water droplet is $ S_{exact} = 0.01\pi $ due to $\nabla \cdot \bm{u}=0$. Fig. \ref{Fig:water-periodic-movement}(b) depicts the convergence of the relative errors $E(area) = |S_{numerical} - S_{exact}|/S_{exact} $ on the uniform mesh with grid refinement, which suggests that nearly second-order accuracy is obtained for both the present and the original method. 

This case is also calculated on a multi-resolution grid where the effective resolution at the finest level is $256 \times 256$ with the maximum level of resolution at $L_{max} = 4$. The multi-resolution representations at different time are depicted in Fig. \ref{Fig:water-periodic-movement}(c)-(d), where the resolutions of three levels can be observed. As shown in Table. \ref{Table:watermovespeedup}, the speedup ratio is reduced by about 0.8 compared to the uniform grid situation when the resolutions are the same. There are two main reasons: First, the extra procedures related to the data structure lead to the decrease of $m$ in Eq. \ref{Eq:speedup_ratio}. Second, as shown in Fig. \ref{Fig:water-periodic-movement}(c)-(d), some single-phase leaves of gas are located at coarser levels where calculations are carried out with a larger time step.

 Let $M$ represents the total mass of the droplet and $\rho_{i}$ denotes the cell-averaged mass density of cell $i$. Assuming that the droplet occupys cells ranging from the first to the $N$th, the relative mass is calculated by
\begin{equation}
\dfrac{M^{n}}{M^{0}}=\dfrac{\sum_{i=1}^{i=N} \alpha^{n} \rho_{i}^{n}}{\sum_{i=1}^{i=N} \alpha^{0} \rho_{i}^{0}},
\end{equation}
where the superscripts 0 and n represent the initial condition and the nth time step, respectively. Fig. \ref{Fig:water-periodic-movement}(a) depicts the relative mass of the water droplet during the computation. It can be observed that the relative mass is always equal to 1, which confirms that the present method has no conservation error.

%
%%%%%%%%%%%%%%%%%%%%%%%%%%%%%%%%%%%%%
\begin{table}
\centering
\begin{tabular}{|c|c|c|c|}
\hline
\multirow{2}*{Resolution}&\multicolumn{2}{c|}{Calculation time ($ s $)}&\multirow{2}*{Speedup ratio $r_s$}\\
\cline{2-3}
&Original&Modified&\\
\hline
\multicolumn{4}{|c|}{Uniform grid - 2D}\\
\hline
$32\times32$&140.585&91.3541&1.539\\
\hline
$64\times64$&618.656&321.757&1.923\\
\hline
$128\times128$&3736.34&1609.05&2.322\\
\hline
$256\times256$&26601.7&9259.04&2.873\\
\hline
\multicolumn{4}{|c|}{Uniform grid - 3D}\\
\hline
$128\times128\times128$&1001450.6&349911.1&2.862\\
\hline
\multicolumn{4}{|c|}{Multi-resolution grid - 2D}\\
\hline
$256\times256$&12309.6&5970.85&2.062\\
\hline
\end{tabular}
\caption{Periodic motion of a water droplet: Speedup ratios for different resolutions. In this paper, for all cases, the speedup performance of the accelerated method is obtained on a desktop with the Intel Core i7-4790 Processor (8 M Cache, 3.60 GHz, 8 cpus, 8GB of RAM). The intel Threading Building Blocks(TBB) library \cite{4636091} is used to map logical tasks to physical threads.}
\label{Table:watermovespeedup}
\end{table}
%%%%%%%%%%%%%%%%%%%%%%%%%%%%%%%%%%%%%%%%%%
%

%
%%%%%%%%%%%%%%%%%%%%%%%%%%%%%%%%%%%
%%%%%%%%%%%%%%%%%%%%%%%%%%%%%%%%%%%
%

\subsection{Static drop with surface tension}
\label{subsec4.3}
In this case, we consider a static liquid drop of radius $R=0.1$ embedded in the air to test the Laplace law. With the surface tension coefficient $\sigma=36$, the initial conditions are given by
\begin{equation}
\left\{
\begin{aligned}
\rho&=1, p_{0}=10^{3}, B=100, \gamma=1.4, \mu = 1.6\times 10^{-5}&\quad air, \\
\rho&=1000, p_{0}=10^{3}, B=5.0\times10^{5}, \gamma=7.15,\mu=10^{-3}&\quad liquid\  drop, \\
\phi&=-0.1+\sqrt{(x-0.5)^{2}+(y-0.5)^{2}}&\quad level\ set.
\end{aligned}
\right.
\end{equation}
The computational domain is a unit square with no-slip boundary conditions, which is discretized by a uniform grid with increasing resolution from $ 32\times32 $ to $ 256\times256 $ as well as a multi-resolution grid where the effective resolution at the finest level is $256 \times 256$. Speedup ratios for different resolutions are given in Table. \ref{Table:static drop speedup}. 

\begin{figure}[p]
\centering
\includegraphics[width=1.0\textwidth]{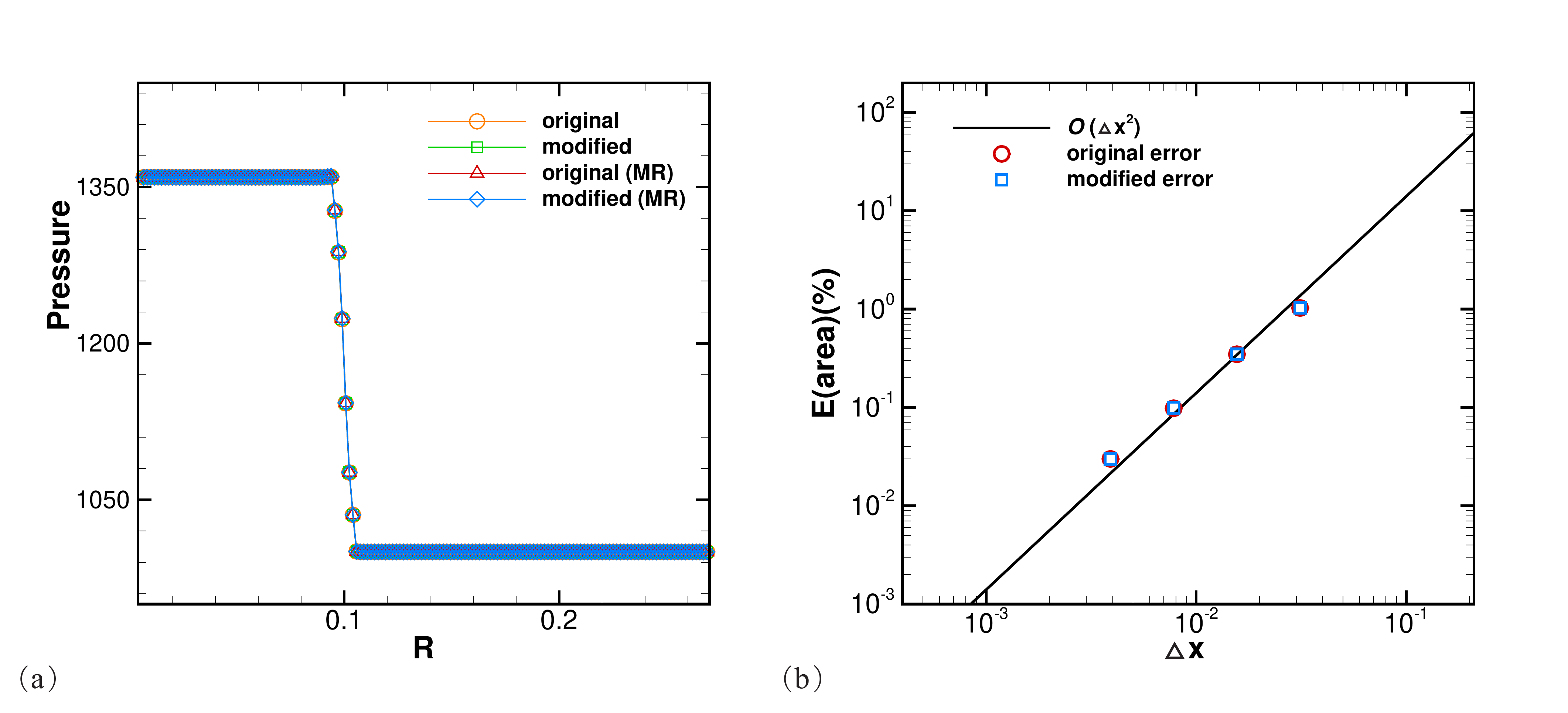}
\caption{Static drop with surface tension:  (a) Pressure distribution; (b) Relative error according to resolution. (a) is obtained on the uniform grid involving $256 \times 256$ cells as well as the multi-resolution grid  with an effective resolution of $256\times256$ at the finest level while (b) is obtained on the uniform mesh with grid refinement.  }
\label{Fig:static_drop}
\end{figure}

Because of the capillary effects, the pressure distribution of steady state should satisfy the Laplace law Eq. (\ref{Eq:surfacetension}). The theoretical pressure jump can be calculated as
$$
\Delta p_{exact}=p_{liquid}-p_{air}= \frac{\sigma}{R}=360,
$$
and the relative error is obtained by
$$
E(\Delta p)= \frac{|\Delta p_{exact}-\Delta p_{numerical}|}{|\Delta p_{exact}|}.
$$
Fig. \ref{Fig:static_drop}(a) shows the steady state pressure distribution obtained at the finest resolution $256 \times 256$. On both the uniform grid and the multi-resolution grid, the results of the accelerated method are almost the same as the original results. From Fig. \ref{Fig:static_drop}(b), it can be observed that nearly second-order accuracy is obtained for both the present and the original method.

%
%%%%%%%%%%%%%%%%%%%%%%%%%%%%%%%%%%%%%%%%%%%
\begin{table}
\centering
\begin{tabular}{|c|c|c|c|}
\hline
\multirow{2}*{Resolution}&\multicolumn{2}{c|}{Calculation time($ s $)}&\multirow{2}*{Speedup ratio $r_s$}\\
\cline{2-3}
&Original&Modified&\\
\hline
\multicolumn{4}{|c|}{Uniform grid}\\
\hline
$32\times32$&230.491&144.468&1.595\\
\hline
$64\times64$&993.682&461.515&2.153\\
\hline
$128\times128$&6200.05&2449.27&2.531\\
\hline
$256\times256$&43680.1&14068.5&3.1048\\
\hline
\multicolumn{4}{|c|}{Multi-resolution grid}\\
\hline
$256\times256$&9441.02&19285.4&2.043\\
\hline
\end{tabular}
\caption{Static drop with surface tension: Speedup ratios for different resolutions.}
\label{Table:static drop speedup}
\end{table}
%%%%%%%%%%%%%%%%%%%%%%%%%%%%%%%%%%%%%%%%
%

%
%%%%%%%%%%%%%%%%%%%%%%%%%%%%%%%%%%%%%%%%%%%
%%%%%%%%%%%%%%%%%%%%%%%%%%%%%%%%%%%%%%%%%%%
%

\subsection{Recovering a circle shape}
\label{subsec4.4}
\begin{figure}[p]
\centering
\includegraphics[width=1.0\textwidth]{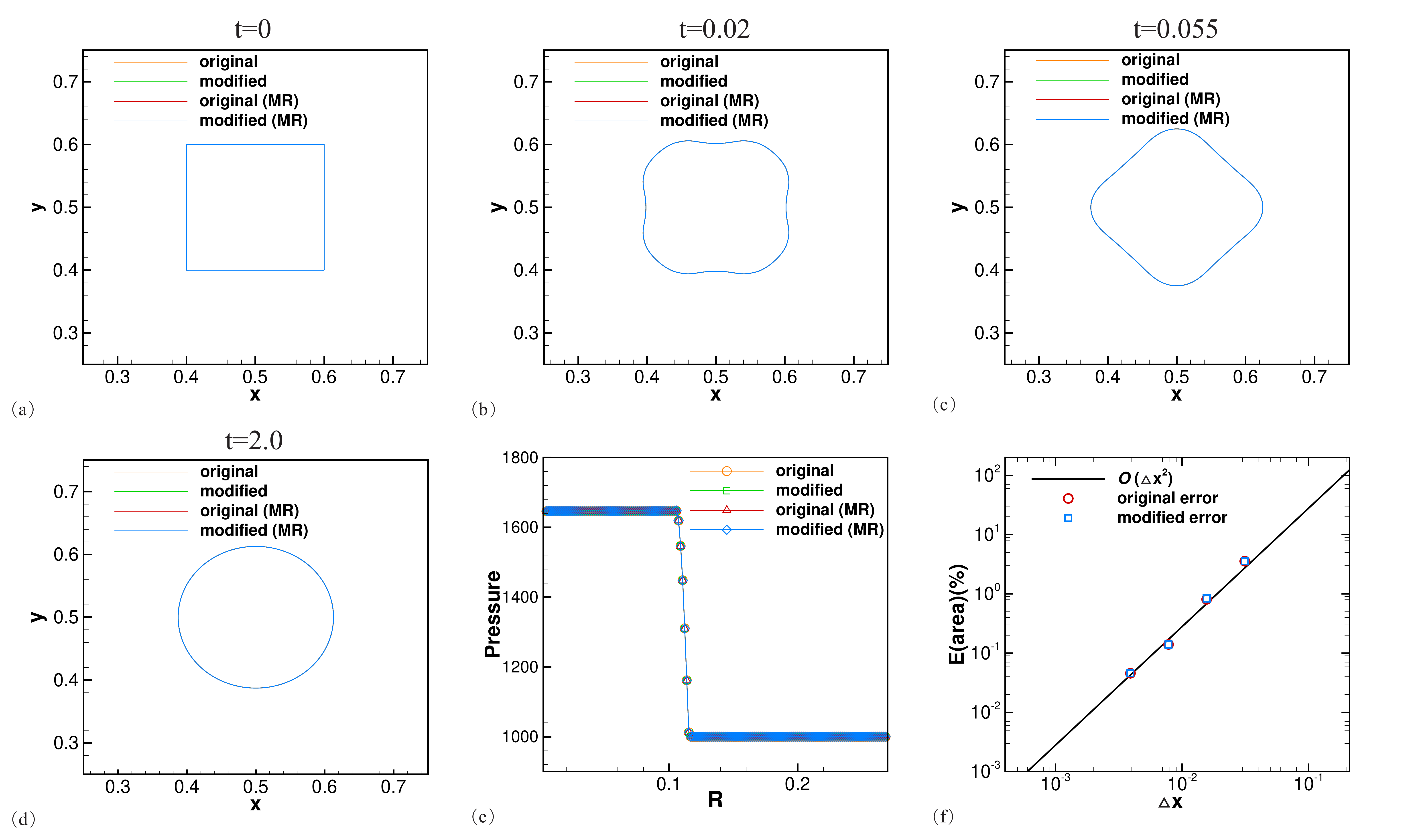}
\caption{Recovering a circle shape:  (a)-(d) The shapes of the droplet at different time; (e) Pressure distribution; (f) Relative error according to resolution. (a)-(e) are obtained on the uniform grid involving $256 \times 256$ cells as well as the multi-resolution grid with an effective resolution of $256\times256$ at the finest level while (f) is obtained on the uniform mesh with grid refinement}
\label{Fig:circle recovering}
\end{figure}

Taken from \cite{schmidmayer2017model}, a square drop deforming in a gas-like fluid is computed for dynamic verification of surface tension. We consider a $0.2 \times 0.2$ square liquid drop placed at the center of a $1 \times 1$ square computational domain with no-slip boundary conditions. With the surface tension coefficient $\sigma=73$, the initial conditions are given by
\begin{equation}
\left\{
\begin{aligned}
\rho&=1, p_{0}=10^{3}, B=100, \gamma=1.4, \mu = 1.6\times 10^{-5}&\quad gas, \\
\rho&=1000, p_{0}=10^{3}, B=5.0\times10^{5}, \gamma=7.15,\mu=10^{-3}&\quad liquid\  drop, \\
\phi&=max(|x-0.5|,|y-0.5|)-0.1&\quad level\ set.
\end{aligned}
\right.
\end{equation}
Note that the level-set function given above is not a strict signed distance function so it has to be reinitialized before the simulation. In this case, the simulations are carried out on the uniform grid with increasing resolution from $ 32\times32 $ to $ 256\times256 $ and the multi-resolution grid where the effective resolution at the finest level is $256 \times 256$. Due to the capillary effects, the shape of the liquid drop will evolve to a circle and the pressure of the steady state will satisfy the Laplace law Eq. (\ref{Eq:surfacetension}). The flow field reaches the steady state at time $t = 2.0$. In Fig. \ref{Fig:circle recovering}(a)-(d), with the same resolution $256\times256$, the interfaces of the droplet obtained by the present method are compared with the original results at different time, which are almost identical.

Unlike Schmidmayer et al. \cite{schmidmayer2017model}, we employ a weakly compressible model in this case so that mass conservation implies volume conservation. With the initial area of the square drop $S_{initial}=0.04$, the radius of the final circle can be calculated by $R = \sqrt{\frac{0.04}{\pi}}=0.112838$. Then the theoretical pressure jump can be obtained by 
$ [\,p\,] = \frac{\sigma}{R}= 646.94566$. Fig. \ref{Fig:circle recovering}(e) depicts the steady state pressure distribution obtained by the present method, which shows a good agreement with the original results. In Fig. \ref{Fig:circle recovering}(f), relative error $E(area)=|S_{final}-S_{initial}|/S_{initial}$ is plotted versus $\Delta x$, which suggests about second-order convergence for both the present and the original method.

\begin{figure}[p]
\centering
\includegraphics[width=1.0\textwidth]{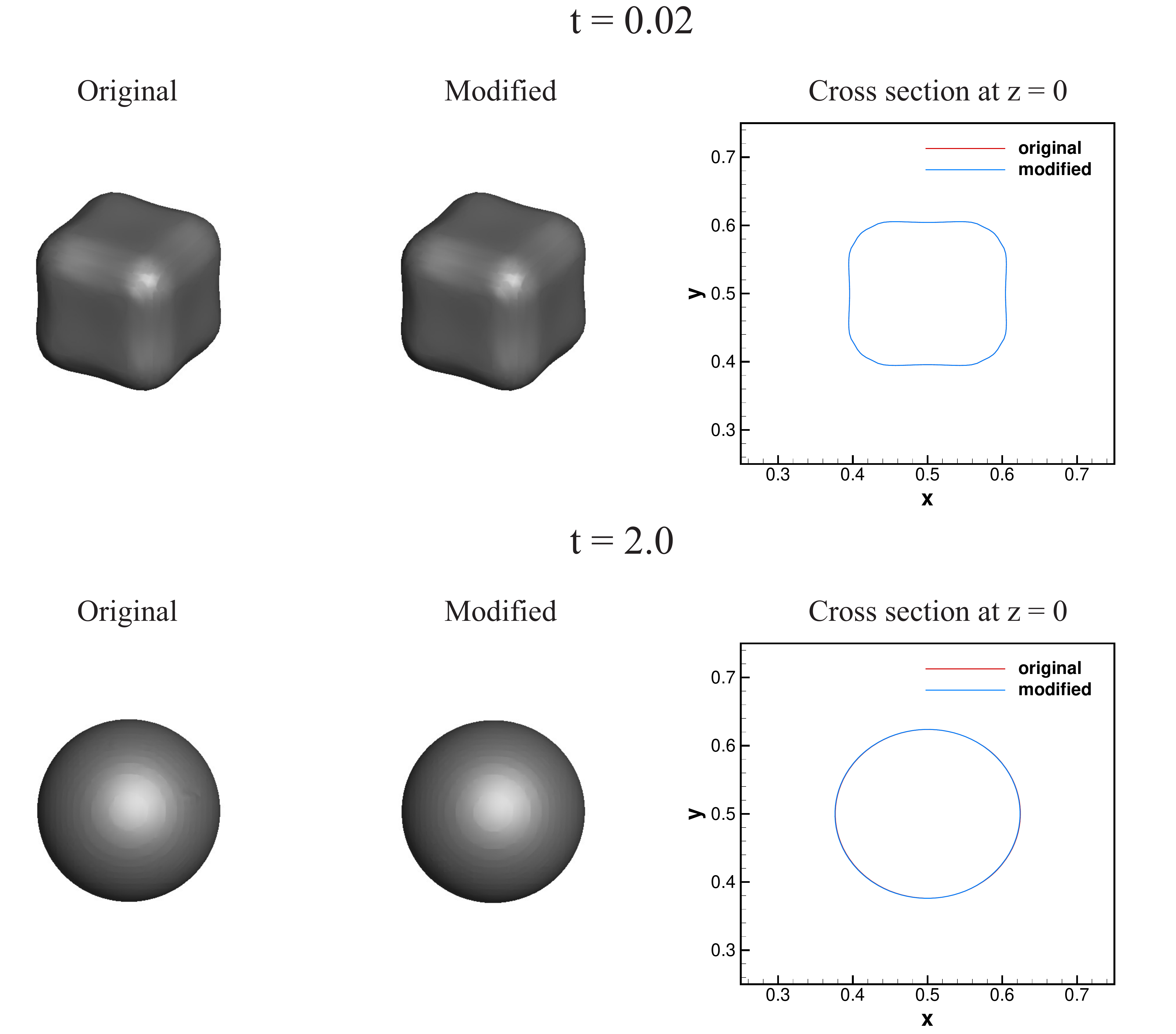}
\caption{Recovering a circle shape (3D): The domain is discritized by a uniform grid involving $128\times128\times128$ cells.}
\label{Fig:circle recovering 3D}
\end{figure}

To repeat this case in three dimensions, the velocity component in the z-axis direction $w$ is set to 0 and the level-set field is initialized by $phi=max(|x-0.5|,|y-0.5|,|z-0.5| )-0.1$. After reinitializing the level-set field, the simulation is carried out on a uniform grid involving $128\times128\times128$ cells. Speedup ratios for different resolutions are listed in Table. \ref{Table:circle recovering speedup}. With the resolution as same as in two dimensions, additional speedup ratio of about 0.4 is obtained in three dimensions. Fig. \ref{Fig:circle recovering 3D} depicts the 3D interfaces of the droplet at different time. At the cross section of $z=0$, we can find that the interfaces obtained by the present method are consistent with those in the original results. Note that the interfaces in Fig. \ref{Fig:circle recovering 3D} are slightly different from those in Fig. \ref{Fig:circle recovering} due to their lower resolution. 

%
%%%%%%%%%%%%%%%%%%%%%%%%%%%%%%%%%%%%%%%%%%%%%%%%%%%%
\begin{table}
\centering
\begin{tabular}{|c|c|c|c|}
\hline
\multirow{2}*{Resolution}&\multicolumn{2}{c|}{Calculation time($ s $)}&\multirow{2}*{Speedup ratio $r_s$}\\
\cline{2-3}
&Original&Modified&\\
\hline
\multicolumn{4}{|c|}{Uniform grid - 2D}\\
\hline
$32\times32$&112.889&71.778&1.573\\
\hline
$64\times64$&604.497&348.205&1.736\\
\hline
$128\times128$&3141.19&1270.66&2.472\\
\hline
$256\times256$&21889.7&7168.9&3.053\\
\hline
\multicolumn{4}{|c|}{Uniform grid - 3D}\\
\hline
$128\times 128 \times128$&946140.3&328314.1&2.882\\
\hline
\multicolumn{4}{|c|}{Multi-resolution grid - 2D}\\
\hline
$256\times256$&9206.3&5174.4&1.779\\
\hline
\end{tabular}
\caption{Recovering a circle shape: Speedup ratios for different resolutions.}
\label{Table:circle recovering speedup}
\end{table}
%%%%%%%%%%%%%%%%%%%%%%%%%%%%%%%%%%%%%%%%%%%%%%
%

%
%%%%%%%%%%%%%%%%%%%%%%%%%%%%%%%%%%%%%%%%%%%%%%
%%%%%%%%%%%%%%%%%%%%%%%%%%%%%%%%%%%%%%%%%%%%%%
%

%
%%%%%%%%%%%%%%%%%%%%%%%%%%%%%%%%%%%%%%%%%%%%%%
%%%%%%%%%%%%%%%%%%%%%%%%%%%%%%%%%%%%%%%%%%%%%%
%

\subsection{Drop oscillation}
\label{subsec4.5}
\begin{figure}[p]
\centering
\includegraphics[width=1.0 \textwidth]{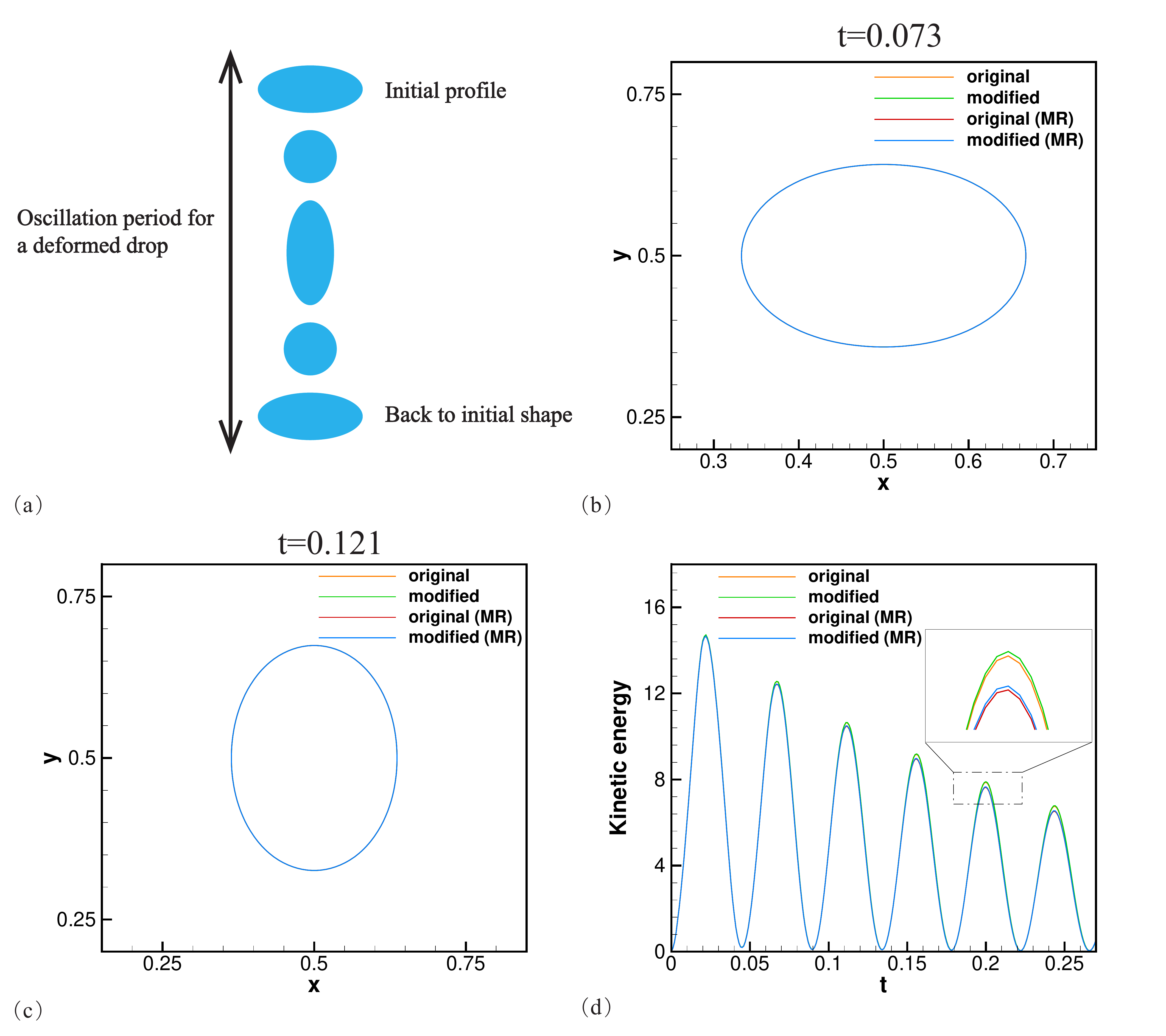}
\caption{Drop oscillation:  (a) oscillation period; (b)-(c) shapes of droplet at different time; (d) Evolution of kinetic energy. The domain is discretized by the uniform grid involving $128\times128$ cells as well as the multi-resolution grid with an effective resolution of $128\times128$ at the finest level.}
\label{Fig:drop oscillation}
\end{figure}

In this case, we consider a $1\times 1$ square computational domain with a ellipsoidal drop located at its center to access the prediction of capillary instabilities \cite{lundgren1988oscillations}. Surrounded by a gas-like fluid, the ellipsoidal drop is initially at rest with zero kinetic energy. Then the existence of surface tension results in the oscillatory behaviour of the liquid drop along with the translation between the potential energy and the kinetic energy. As shown in Fig. \ref{Fig:drop oscillation}(a), in on cycle, the shape of the liquid interface changes from an ellipse to a circle, and then it deforms into an ellipse again. With the surface tension coefficient $\sigma=341.642$, the initial conditions are give as
\begin{equation}
\left\{
\begin{aligned}
\rho&=1, p_{0}=10^{4}, B=500, \gamma=1.4&\quad gas, \\
\rho&=100, p_{0}=10^{4}, B=2.0\times 10^{5}, \gamma=7.15&\quad liquid\  drop, \\
\phi&=-1.0+\sqrt{(x-0.5)^{2}/ 0.2^{2}+(y-0.5)^{2}/ 0.12^{2}}&\quad level\ set.
\end{aligned}
\right.
\label{Eq:elliptic equation}
\end{equation}
 Note that the level-set function $\phi$ has to be reinitialized before the simulation because the elliptic equation given in Eq. (\ref{Eq:elliptic equation}) is not a signed distance function. This case is calculated on the uniform grid involving $128 \times 128$ cells and the multi-resolution grid with the effective resolution $128 \times 128$ at the finest level. The speedup ratio obtained on the uniform grid is $s = 2.196$ and decreases to $s = 1.864$ on the multi-resolution grid. 

The shapes of the droplet at different time are given in Fig. \ref{Fig:drop oscillation}(b)-(c), which show a good agreement between the results obtained by the present method and the orignal method. Fig. \ref{Fig:drop oscillation}(d) depicts the evolution of the kinetic energy versus simulation time. It can be observed that the peaks of the kinetic energy obtained on the multi-resolution grid are lower than those obtained on the uniform grid due to the extra errors generated by the MR algorithm. However, as long as the simulations are carried out on the same grid, the results of the present method are almost identical to the original results. According to the Rayleigh formula which has been extended to two-phase flows \cite{fyfe1988surface}, the oscillation frequency can be obtained by
\begin{equation}
\omega^2=(o^3-o)\frac{\sigma}{(\rho_l+\rho_g)R^3},\quad\quad\quad T=\frac{2\pi}{\omega}.
\label{Eq:oscillation}
\end{equation}
Here $\rho_l$ and $\rho_g$ are the density of the liquid drop and the surrounding gas-like fluid, respectively. In addition, $o$ is the oscillation mode and $R$ is the radius of the liquid drop at the equilibrium state. Following \cite{perigaud2005compressible,luo2015conservative}, we choose $o=2$ and $R=0.15825$ so that the theoretical value of the oscillation period is $T=0.0878$. Evaluated from Fig. \ref{Fig:drop oscillation}(d), on both the uniform gird and the multi-resolution grid, the oscillation period obtained by the present method is $T = 0.0881$ as same as in the original results. The error is about $0.3\%$ as the same in \cite{luo2015conservative}. 
%
%%%%%%%%%%%%%%%%%%%%%%%%%%%%%%%%%%%%%%%%%%%%%
%%%%%%%%%%%%%%%%%%%%%%%%%%%%%%%%%%%%%%%%%%%%%
%

\subsection{Liquid filament contraction}
\label{subsec4.6}
\begin{figure}[p]
\centering
\includegraphics[width=0.6\textwidth]{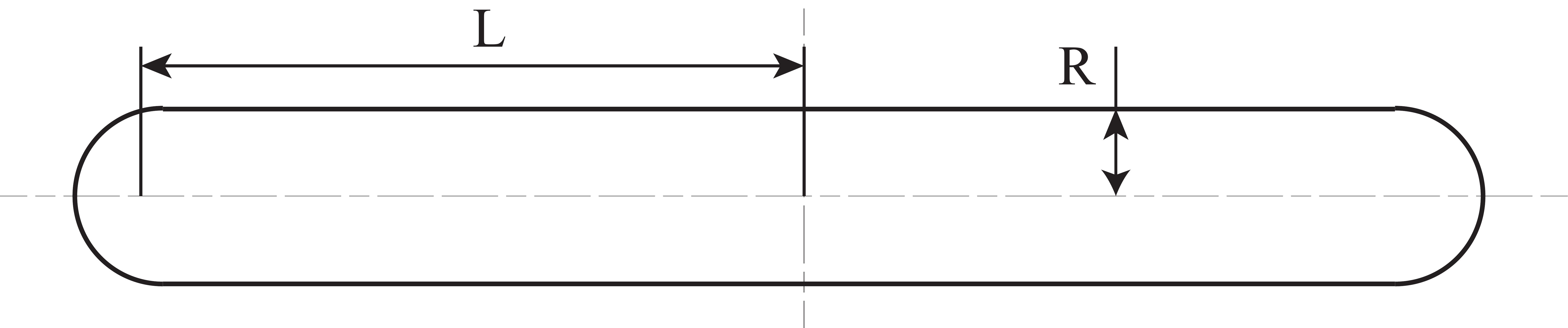}
\caption{Liquid filament contraction: Geometry of a contracting filament}
\label{Fig:geometry filament}
\end{figure}

Proposed by Schulkes et al. \cite{Schulkes1996contraction}, the liquid filament contraction is considered to validate the present method for problems with coupled effects of viscosity and surface tension. As shown in Fig. \ref{Fig:geometry filament}, the initial shape of the filament is a cylinder with two hemispherical caps at both of its ends. In addition, the half-length of the filament is $L$ and the radius of the cylinder as well as the hemispherical is $R$. Two important dimensionless parameters for this problem are the aspect ratio of the filament $L_0 = L/R$ and the Ohnesorge number $Oh=\mu/\sqrt{\rho \sigma R}$, which measures the relative importance of the viscous force to the surface tension force. In this case, we choose $L_0 = 9.0$ and $Oh = 8.6 \times 10^{-3}$ so that the filament will contract into a single drop. With the the surface tension coefficient $\sigma=72\ mNm^{-1}$, the initial conditions are given as
\begin{equation}
\left\{
\begin{aligned}
\rho&=1\ kg/m^{3}, p_{0}=10^{4}Pa, B=5.0 \times10^2 Pa,\quad \text{air}\\
 \gamma&=1.4, \mu = 1.6\times 10^{-5}Pa\ s    \\
\rho&=1000\ kg/m^{3}, p_{0}=10^{4}Pa, B=3.0 \times10^6 Pa,\quad \text{liquid\ filament}\\
 \gamma&=7.15, \mu = 1\times 10^{-3}Pa\ s    \\ 
\phi&=\left\{
\begin{array}
{ll}{-0.19+|x| \ mm} & {z<1.52} \\ 
{-0.19+\sqrt{x^{2}+(z-1.52)^{2}} \ mm} & {z \geq 1.52}
\end{array}
\right. \quad \text { level-set. }
\end{aligned}
\right.
\end{equation}
Due to the symmetry property of this case, the computational domain is $1/4$ of the physical domain with 1.0 mm in the x-axis direction and 2.0 mm in the z-axis direction. The computations are carried out on the uniform grid with increasing resolutions from $16\times32$ to $128 \times 256$ and the multi-resolution grid where the effective resolution at the finest level is $128 \times 256$. To increase the computational accuracy, a two dimensional axisymmetric model is used in this case. 

\begin{figure}[p]
\centering
\includegraphics[width=1.0\textwidth]{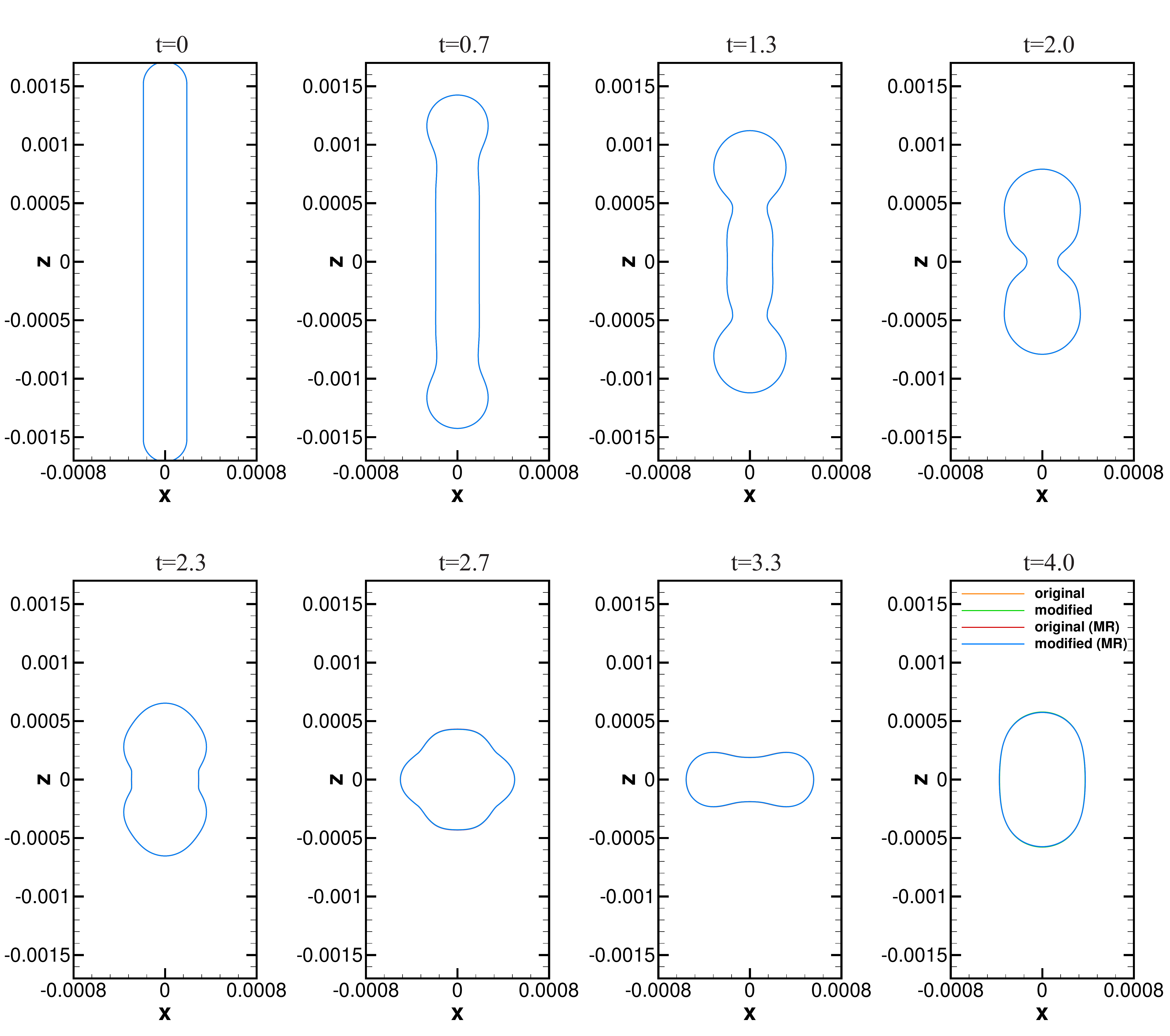}
\caption{Liquid filament contraction (2D): Evolution of the contracting liquid filament. The computational domain is $1/4$ of the physical domain with 1.0 mm in the x-axis direction and 2.0 mm in the z-axis direction, which is discretized by the uniform grid involving $128 \times 256$ cells as well as the multi-resolution grid with an effective resolution of $128\times256$ at the finest level . }
\label{Fig:evolution filament}
\end{figure}

Fig. \ref{Fig:evolution filament} depicts the evolution of the contracting liquid filament at the finest resolution $128 \times 256$. The filament contracts quickly and forms bulbous ends which become larger and larger during the evolution. The filament fails to pinch off its bulbous ends because the capillary forces are not strong enough to overcome the viscous force. After a series of oscillations with diminishing amplitudes, the filament finally become a sphere (not shown here). It can be observed that during the evolution the shapes of the filament obtained by the present method are consistent with the original results on both the uniform grid and the multi-resolution grid.

\begin{figure}[p]
\centering
\includegraphics[width=1.0\textwidth]{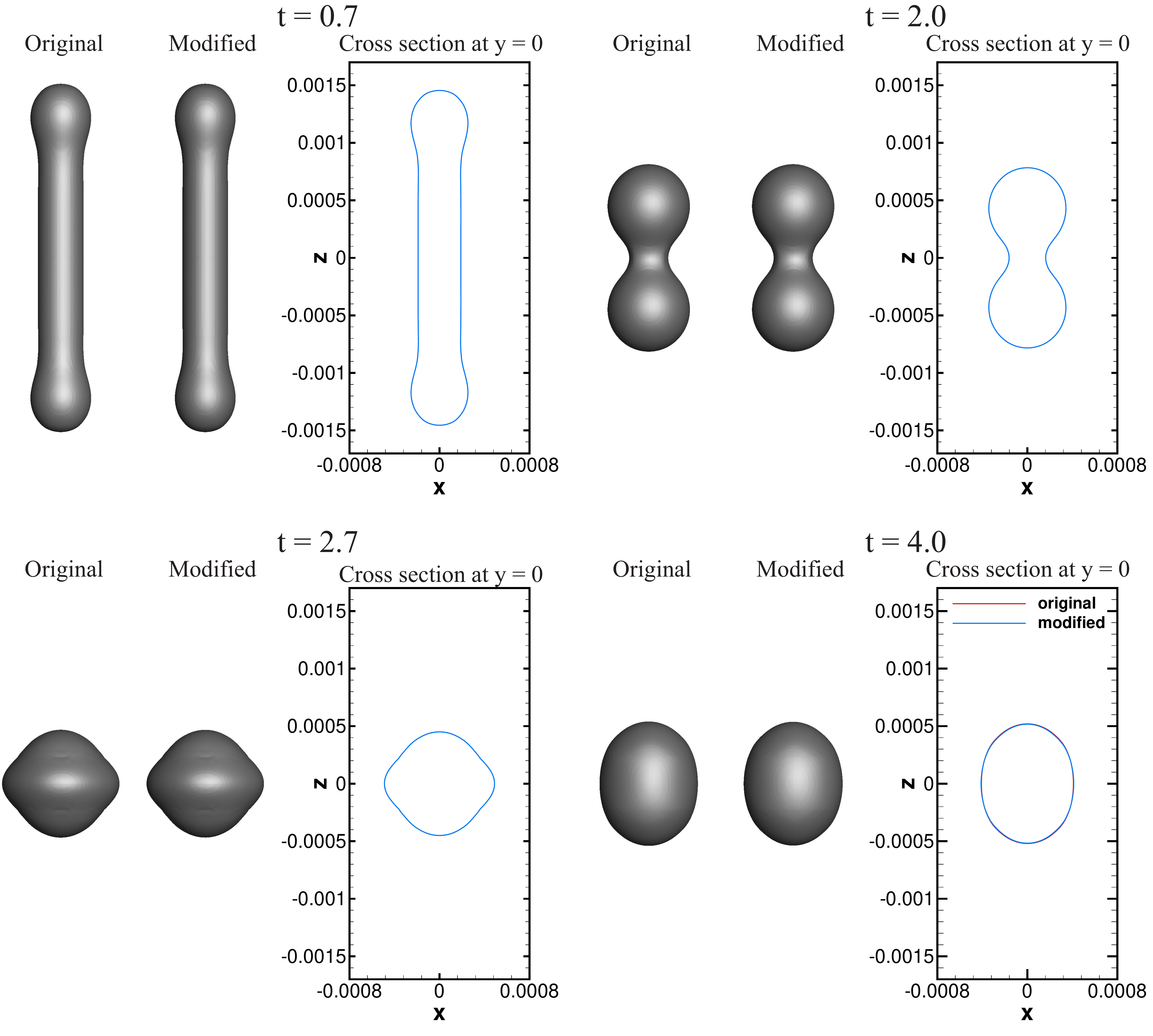}
\caption{Liquid filament contraction (3D): Evolution of the contracting liquid filament. The computational domain is $1/8$ of the physical domain with 1.0 mm in x-axis direction, 1.0 mm in y-axis direction and 2.0 mm in z-axis direction, which is discretized by a uniform grid involving $128 \times 128 \times 256$ cells.}
\label{Fig:evolution filament (3D)}
\end{figure}

To calculate the 3D version of this case, we modify the initial conditions by setting the velocity component in the z-axis direction to 0 and initializing the level-set field by
\begin{equation}
\phi=\left\{
\begin{array}
{ll}{-0.19+\sqrt{x^{2} +y^{2}} \ mm} & {z<1.52}\  \\ 
{-0.19+\sqrt{x^{2} +y^{2}+(z-1.52)^{2}} \ mm} & {z \geq 1.52}\ 
\end{array}
\right. \nonumber
\end{equation}
Involving $64\times64\times128$ computational cells, the computational domain is $1/8$ of the physical domain with 1.0 mm in x-axis direction, 1.0 mm in the y-axis direction and 2.0 mm in the z-axis direction. Speedup ratios for different resolutions are given in Table. \ref{Table:Liquid filament contraction speedup}. With the resolution as same as in two dimensions, additional speedup ratio of about 0.4 is obtained in three dimensions. The 3D interfaces of the filament obtained by two methods at different time are shown in Fig. \ref{Fig:evolution filament (3D)}. For a better comparison, at the cross section of $y=0$, we depict the interfaces of the filament obtained by the accelerated method and the original method, which are well matched. Note that the interfaces in Fig. \ref{Fig:evolution filament (3D)} are slightly different from those in Fig. \ref{Fig:evolution filament} because their resolutions are lower.

%
%%%%%%%%%%%%%%%%%%%%%%%%%%%%%%%%
\begin{table}
\centering
\begin{tabular}{|c|c|c|c|}
\hline
\multirow{2}*{Resolution}&\multicolumn{2}{c|}{Calculation time($ s $)}&\multirow{2}*{Speedup ratio $r_s$}\\
\cline{2-3}
&Original&Modified&\\
\hline
\multicolumn{4}{|c|}{Uniform grid - 2D}\\
\hline
$16\times32$&236.415&175.885&1.344\\
\hline
$32\times64$&930.915&230.498&1.661\\
\hline
$64\times128$&4325.97&2116.39&2.044\\
\hline
$128\times256$&28849.9&11673.9&2.471\\
\hline
\multicolumn{4}{|c|}{Uniform grid - 3D}\\
\hline
$ 64 \times 64 \times128$&668940.1&275521.6&2.428\\
\hline
\multicolumn{4}{|c|}{Multi-resolution grid - 2D}\\
\hline
$ 128\times256$&17085.6&9735.21&1.755\\
\hline
\end{tabular}
\caption{Liquid filament contraction: Speedup ratios for different resolutions}
\label{Table:Liquid filament contraction speedup}
\end{table}
%%%%%%%%%%%%%%%%%%%%%%%%%%%%%%%%%%%%
%

%
%%%%%%%%%%%%%%%%%%%%%%%%%%%%%%%%%%%%
%%%%%%%%%%%%%%%%%%%%%%%%%%%%%%%%%%%%
%

%
%%%%%%%%%%%%%%%%%%%%%%%%%%%%%%%%%%%%%%%%%
%%%%%%%%%%%%%%%%%%%%%%%%%%%%%%%%%%%%%%%%%
%

%
%%%%%%%%%%%%%%%%%%%%%%%%%%%%%%%%%%%%%%%%%
%%%%%%%%%%%%%%%%%%%%%%%%%%%%%%%%%%%%%%%%%
%

\subsection{Equilibrium shape of a water drop resting on a wall}
\label{subsec4.7}
\begin{figure}[p]
\centering
\includegraphics[width=1.0\textwidth]{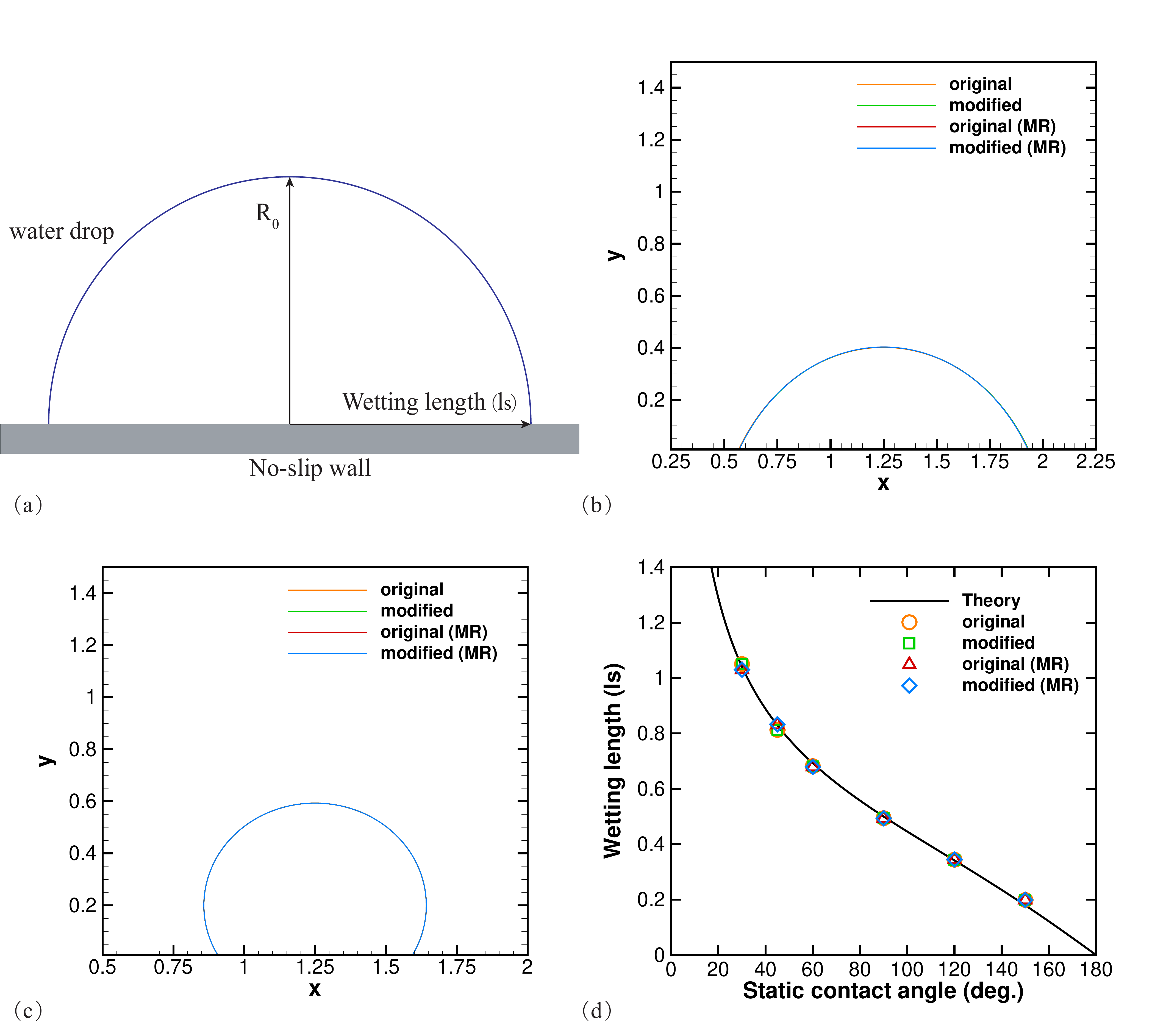}
\caption{Equilibrium shape of a water drop resting on a wall: (a) Setup for the drop spreading; (b) Equilibrium shape of the water drop with the static contact angle $\theta_s = 60\degree$; (c) Equilibrium shape of the water drop with the static contact angle $\theta_s = 120\degree$; (d) Comparison of wetting length($l_s$) with the theory. The domain is discretized by the uniform grid involving $128\times128$ cells as well as the multi-resolution grid with an effective resolution of $128\times128$ at the finest level .}
\label{Fig:contact line}
\end{figure}

By employing a curvature boundary condition proposed by Luo et al. \cite{luo2016curvature}, we consider an equilibrium water drop on a horizontal wall to validate the present method for problems with a static contact angle. As illustrated in Fig. \ref{Fig:contact line}(a), the initial shape of the water drop is a semi-circle with the radius $R_0=0.5$, which initially rests on the wall with the contact angle $90\degree$. Due the the difference between the transient contact angle and the static contact angle, the water drop will deform and its final shape is determined by two parameters, the static contact angle and the $E\ddot{o}tv\ddot{o}s$ number $Eo = \rho g R_0^2 / \sigma$. In this case, we ignore the gravity so that $Eo = 0$ and the equilibrium shape of the water drop is a circular cap determined by the static angle only. The computational domain is a $2.5\times 2.5$ square with no-slip boundary conditions. With the surface tension coefficient $\sigma=36$, the initial conditions are given by
\begin{equation}
\left\{
\begin{aligned}
\rho&=1, p_{0}=10^{3}, B=100, \gamma=1.4, \mu = 4\times 10^{-3}&\quad air, \\
\rho&=1000, p_{0}=10^{3}, B=5.0\times10^{5}, \gamma=7.15,\mu=0.25&\quad liquid\  drop, \\
\phi&=\sqrt{(x-1.25)^2+y^2}-0.5&\quad level\ set.
\end{aligned}
\right.
\end{equation}
With different static contact angles, this case is calculated on the uniform grid involving $128 \times 128$ cells and the multi-resolution grid where the effective resolution at the finest level is $128 \times 128$. When the water drop reaches the steady state, the wetting length can be calculated by
\begin{equation}
l_s = R_0\sin{\theta_s}\sqrt{\frac{\pi /2}{\theta_s - \sin{\theta_s}\cos{\theta_s}}}
\end{equation}
according to the volume conservation. Fig. \ref{Fig:contact line}(b)-(c) show the equilibrium shapes of the water drop with different static contact angles. It can be observed that the interfaces obtained by the accelerated method are almost identical to those obtained by the original method. The numerical results of the wetting length are compared with the theoretical values in Fig. \ref{Fig:contact line}(d). It is found that the simulation results of both the present and the original method agree with the theoretical values quite well. Speedup ratios for different static contact angles are listed in Table. \ref{Table:contactline speedup}.

%
%%%%%%%%%%%%%%%%%%%%%%%%%%%%%%%%%%%%%%%
%%%%%%%%%%%%%%%%%%%%%%%%%%%%%%%%%%%%%%%
%

%
%%%%%%%%%%%%%%%%%%%%%%%%%%%%%%%%%%%%%%%%
\begin{table}
\centering
\begin{tabular}{|c|c|c|c|c|c|c|}
\hline
Static contact angle($\theta_s$)&$30\degree$&$45\degree$&$60\degree$&$90\degree$&$120\degree$&$150\degree$\\
\hline
\multicolumn{7}{|c|}{Uniform grid}\\
\hline
Speedup ratio&2.171&2.078&2.039&2.0&2.162&2.139\\
\hline
\multicolumn{7}{|c|}{Multi-resolution grid}\\
\hline
Speedup ratio&1.714&1.724&1.726&1.717&1.718&1.723\\
\hline
\end{tabular}
\caption{Equilibrium shape of a water drop resting on a wall: Speedup ratios for different static contact angles}
\label{Table:contactline speedup}
\end{table}
%%%%%%%%%%%%%%%%%%%%%%%%%%%%%%%%%%%%%%%
%

\section{Concluding remarks}
\label{sec5}
In this paper, we have developed an accelerated sharp-interface method for compressible and weakly compressible multiphase flows. This method inherits the simplicity of the original GFM-like methods of Hu et al. \cite{hu2006conservative} and Luo et al. \cite{luo2015conservative}. Different from the semi-implicit methods \cite{Kwatra2009Method,Kadioglu2008Adaptive,Yabe2007Unified} and the free-surface conditions \cite{Carrica2010Unsteady, Kim2003FREESURFACE}, the present method will not generate extra oscillations and makes no assumptions for the gas phase (or the less viscous and dense phase).  The conservative property of the original methods is violated due to the separated evolution of two fluids, which is handled by the interfacial flux correction. Furthermore, the present method offers a fairly direct way to be combined with the wavelet-based adaptive multi-resolution algorithm \cite{han2014adaptive}. A number of numerical examples are calculated on both the uniform grid and the multi-resolution grid to validate the accelerated method. It is shown that the simulations with the present method is in good agreement with the original results. The higher the grid resolution is, the greater the acceleration performance we obtain. When the number of grid points is around $10000$, for all cases, the computational time of our accelerated method is only about $1/2$ of the original one. In future work we plan to extend the accelerated method to handle the solid-fluid coupling problems as the time step of fluid is usually much larger than the time step of solid.

%\appendix

%\section{Section in Appendix}
%\label{appendix-sec1}

%Sample text. Sample text. Sample text. Sample text. Sample text. Sample text. 
%Sample text. Sample text. Sample text. Sample text. Sample text. Sample text. 
%Sample text. 

%% References


%\bibliographystyle{elsarticle-num}
% \bibliographystyle{elsarticle-harv}
% \bibliographystyle{elsarticle-num-names}
% \bibliographystyle{model1a-num-names}
% \bibliographystyle{model1b-num-names}
% \bibliographystyle{model1c-num-names}
% \bibliographystyle{model2-names}
% \bibliographystyle{model3a-num-names}
% \bibliographystyle{model3-num-names}
% \bibliographystyle{model4-names}
% \bibliographystyle{model5-names}
% \bibliographystyle{model6-num-names}

\bibliographystyle{plain}
\bibliography{ref}
\end{document}